# Heuristics as conceptual lens for understanding and studying the usage of bibliometrics in research evaluation

Lutz Bornmann[1] · Julian N. Marewski[2]

## Abstract
While bibliometrics are widely used for research evaluation purposes, a common theoretical framework for conceptually understanding, empirically studying, and effectively teaching its usage is lacking. In this paper, we outline such a framework: the fast-and-frugal heuristics research program, proposed originally in the context of the cognitive and decision sciences, lends itself particularly well for understanding and investigating the usage of bibliometrics in research evaluations. Such evaluations represent judgments under uncertainty in which typically not all possible options, their consequences, and those consequences' probabilities of occurring may be known. In these situations of incomplete information, candidate descriptive and prescriptive models of human behavior are heuristics. Heuristics are simple strategies that, by exploiting the structure of environments, can aid people to make smart decisions. Relying on heuristics does not mean trading off accuracy against effort: while reducing complexity, heuristics can yield better decisions than more information-greedy procedures in many decision environments. The prescriptive power of heuristics is documented in a cross-disciplinary literature, cutting across medicine, crime, business, sports, and other domains. We outline the fast-and-frugal heuristics research program, provide examples of past empirical work on heuristics outside the field of bibliometrics, explain why heuristics may be especially suitable for studying the usage of bibliometrics, and propose a corresponding conceptual framework.

✉ Lutz Bornmann
 bornmann@gv.mpg.de

 Julian N. Marewski
 julian.marewski@unil.ch

[1] Division for Science and Innovation Studies, Administrative Headquarters of the Max Planck Society, Hofgartenstr. 8, 80539 Munich, Germany

[2] Faculty of Business and Economics, Université de Lausanne, Quartier UNIL-Dorigny, Bâtiment Internef, 1015 Lausanne, Switzerland

## Introduction

Imagine the following scene: Agonizing with burning pain in his chest, a man is brought into a hospital. The medical personnel must make a decision: does the man's state warrant speedy transfer to the coronary care unit or would he be better given a normal hospital bed for further observation? The responsible doctor does not need much time for reflection; that doctor refers the patient to the colleagues at the coronary care unit. In describing a medical situation like this one, Gigerenzer (e.g., 2007) asks how physicians arrive at such decisions. There are at least three ways (see e.g., Gigerenzer 2007; Gigerenzer and Gaissmaier 2011; Marewski and Gigerenzer 2012; Marewski et al. 2010a): First, a doctor might not make a decision in the first place, but simply follow the rule of thumb to 'play it safe' by transferring all patients directly to the coronary care unit. A possible negative outcome of this behavioral option: Many patients, who do not warrant that type of special attention, will end up in the unit, eventually risking to congest it and potentially provoking patient management problems. Related possible negative consequences include exposing patients, perhaps without strict necessity, to health risks (infections) as well as monetary costs incurred by the hospital and/or health insurances. Second, rather than following a cautionary rule of thumb to pass on patients to the coronary care unit, a doctor might rely on a more complex decisional procedure, namely the Heart Disease Predictive Instrument. This second option comes in form of a fairly complicated decision tool: in using the tool, doctors essentially rely on a logistic regression to estimate the probability that a patient better be assigned to the coronary care unit. Third, a doctor might resort to less complex decision aids, for example the decision tree shown in Fig. 1. This tree is built around three questions and can be summarized in terms of three simple if-then rules for decision making: (1) If the electrocardiogram reveals a change in the so-called ST-segment, "the patient is immediately sent to the coronary care unit… [without that any] other information is considered" (Marewski and Gigerenzer 2012, p. 78). (2) "If there is no… [such change, and the chief] complaint is [not] chest pain…, the patient is… [placed into] a regular nursing bed" (Marewski and Gigerenzer 2012, p. 78). (3) If there are no ST-segment changes and if chest pain is the chief complaint and any one of five other signs is there, the patient is directed to the coronary care unit; else the patient ends up in a regular nursing bed. As it turns out, basing an action on the answers to the tree's three questions can lead to better outcomes (e.g., adequate patient assignments) than adopting the more complex regression-based Heart Disease Predictive Instrument (e.g., Gigerenzer 2007).[1]

The decision tree depicted in Fig. 1 is a *fast-and-frugal heuristic* (Gigerenzer et al. 1999). The word "heuristic" has Greek roots and means "serving to find out or discover" (Gigerenzer and Brighton 2009, p. 108). A fast-and-frugal heuristic is a computationally simple decision strategy: a heuristic can base decisions on little information, such as on only a few predictor variables (hence dubbed *frugal*), and can, in doing so, allow making speedy decisions (hence *fast*; see e.g., Gigerenzer and Goldstein 1996; Gigerenzer and Gaissmaier 2011; Hafenbrädl et al. 2016).

---

[1] This introductory example is adapted from Gigerenzer (e.g., 2007; Gigerenzer and Gaissmaier 2011; Marewski and Gigerenzer 2012; Marewski et al. 2010a). While we re-use this example in a context-free way for illustrative purposes, the actual story, a summary of data, and other details can be found in Gigerenzer (e.g., 2007), who describes how the decision tree has been developed by Green and Mehr (1997). Readers familiar with the literature on heuristics—the topic of this article—will know this example from different papers on heuristics (see also Gigerenzer et al. 1999 for a related example). It is an example that aids to illustrate key points on heuristics. In writing this article, we draw on such commonly-used, illustrative and intuitive examples, hoping that they aid those readers unfamiliar with the literature on heuristics to gain an easy intellectual access to that literature and to our arguments. Readers who are familiar with that literature are invited to skip those examples and other contents of this essay that merely serve to introduce research on heuristics.

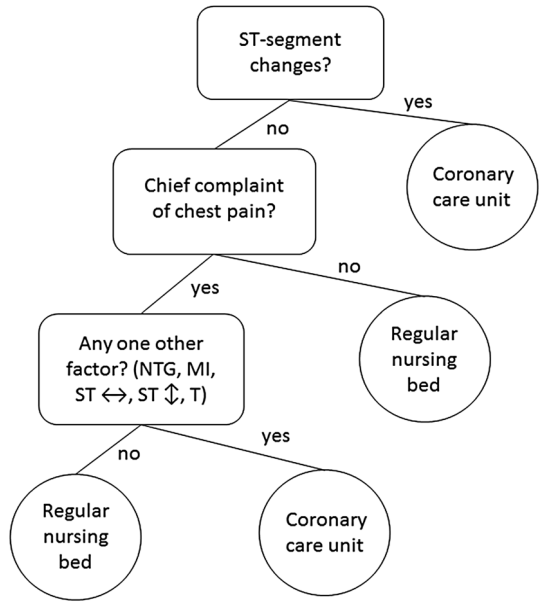

**Fig. 1** "A simple heuristic for deciding whether a patient should be assigned to the coronary care unit or to a regular nursing bed" (Marewski and Gigerenzer 2012, p. 78; *NTG* = nitroglycerin, *MI* = myocardial infarction, *T* = T-waves with peaking or inversion). *Source*: Adapted from Marewski and Gigerenzer (2012, p. 78) and Gigerenzer (2007, p. 174). The figure itself is based on Green and Mehr (1997)

Under certain environmental circumstances, basing decisions on little information can also boost those decisions' accuracy. To explain, in many real-world situations relevant information comes with irrelevant, potentially misleading bits, including mere random noise. In addition to not drawing from irrelevant information, fast-and-frugal heuristics can exploit the relevant, meaningful statistical structures of decision making environments, such as the ways in which decisional options are distributed, or how predictor variables correlate with each other (e.g., Gigerenzer and Brighton 2009; Rieskamp and Dieckmann 2012). In short, as studies in medicine, business, crime, sports, and other domains suggest, fast-and-frugal heuristics can yield smart decisions in classification, forecasting, selection, and other tasks (see Gigerenzer 2008a; Gigerenzer and Gaissmaier 2011; Goldstein and Gigerenzer 2009; Hafenbrädl et al. 2016, for overviews). In addition, heuristics can capture how humans actually make decisions; that is, they can be good descriptive models of behavior (for overviews, see e.g., Gigerenzer and Gaissmaier 2011; Marewski et al. 2010a; but see e.g., Hilbig 2010; Marewski et al. 2018, for discussions).

In line with Hafenbrädl et al. (2016), we believe it can be useful to "actively select a … *conceptual lens*" (p. 216; Italics added) when examining and possibly even shaping a matter. This paper aims to introduce one such conceptual framework—the *fast-and-frugal research program* (e.g., Gigerenzer et al. 1999, 2011)—for studying the application of bibliometrics to research evaluation.[2] Such a framework can aid uncovering which

---

[2] The fast-and-frugal heuristics framework has been developed, starting towards the end of the past century, in the decision sciences (see Hoffrage et al. 2018 for a historical overview). Its founding mothers and fathers were a group of researchers around Gerd Gigerenzer, the so-called *ABC* (= adaptive behavior and cognition) *Research Group*. In introducing and summarizing that framework, this article draws on and borrows from other (overview) pieces on fast-and-frugal heuristics and related topics, among others, Gigerenzer (2007, 2008a, b, 2014), Gigerenzer et al. (1999), Gigerenzer and Brighton (2009), Gigerenzer and Gaissmaier (2011), Gigerenzer and Marewski (2015), Goldstein and Gigerenzer (2009), Hafenbrädl et al. (2016), Marewski et al. (2010a, d), Marewski and Bornmann (2019), Marewski and Gigerenzer (2012), Mousavi and Gigerenzer (2014, 2017), and Todd and Gigerenzer (2000).

*bibliometrics-based heuristics* (henceforth: *BBHs*) are used, when should they be used (and when not), as well as how corresponding heuristics can be investigated. This paper is not intended to advocate bibliometrics as means for research evaluation; the use of bibliometrics for evaluating research is highly debatable (e.g., Gigerenzer and Marewski 2015; Gingras 2016; Osterloh and Frey 2015; MacRobert and MacRoberts 2017). Rather, we outline the fast-and-frugal research program, provide examples of past research on heuristics outside the field of bibliometrics, explain why heuristics might be especially suitable for studying the usage of bibliometrics, and sketch out a corresponding conceptual framework.

## How the study of heuristics can inform the study of bibliometrics

Our central thesis is as follows. While many real-world environments appear to afford making good decisions based on little information (e.g., Gigerenzer and Brighton 2009; Gigerenzer and Gaissmaier 2011; Hafenbrädl et al. 2016), there are still domains in which the dominant line of thought prescribes searching for complex solutions: When decision makers in science (scientists or managers) evaluate units (e.g., research groups or institutions), complex solutions are generally preferred (e.g., because scientific quality is conceived of as a multi-dimensional phenomenon). In research evaluation, complexity can come in at least two different disguises:

First, complexity can arrive at the evaluation stage when many indicators are considered—be they separately presented in long lists or combined into a (weighted) composite indicator. One example is *U-Multirank*,[3] an international university ranking developed recently (van Vught and Ziegele 2012). Alongside the ranking, an indicators book was published explaining the many included indicators on more than 100 pages (U-Multirank 2017). Another example is the *Altmetric Attention Score* for measuring public attention to research which is based on 15 weighted alternative metrics (altmetrics), such as Twitter and Facebook counts.[4]

Second, complexity can come in the disguise of evaluation procedures that include numerous internal and external reviews. Those reviews serve to assess institutions and scientists on multiple evaluation criteria, all of which are then often integrated into a comprehensive joint score or assessment. For example, many universities have established evaluation procedures that start with an internal evaluation in which a research unit (e.g., a university department or faculty) evaluates itself resulting in a self-evaluation report including comprehensive statistics (e.g., Rothenfluh and Daniel 2009). What follows is an external evaluation in which a group of well-known experts visits the department for some days. The visit typically finishes with a report on the unit's strengths and weaknesses. The experts' assessments are based on their own impressions and on the self-evaluation report (Bornmann et al. 2006). For these procedures, the unit and a group of reputable peers are absorbed—leading to longer periods with little active research.

Bibliometrics offer a method for assessing research activities or entities that is frequently employed in research evaluation procedures (in addition to peer review). With bibliometrics, research activities or entities are evaluated by reducing the information considered to publication and citation numbers. Despite its frequent use, the method is often criticized. According to Macilwain (2013), a fundamental defect of quantitative research-assessment tools, such as bibliometrics, is that they are "largely built on sand" (p. 255), which means that they cannot directly measure the 'quality of research' and instead use "weak surrogates, such as the citation indices of individuals" (p. 255). An overview of other critical points on bibliometrics (especially citation analysis) can be found in the paper *The mismeasure of science: citation analysis*, published by MacRoberts and MacRoberts (2017). Critics mention, for example with respect to citation counts, that important publications for the advancement of science are not always highly cited, that researchers create citation circles which cite each other by courtesy, that citations can actually have different purposes and meanings, ranging from endorsement to critique or that citation styles differ (e.g., Adler et al. 2009; Gigerenzer and Marewski 2015; Gläser and Laudel 2007; Osterloh and Frey 2015).

---

[3] See https://www.umultirank.org.

[4] See https://help.altmetric.com/support/solutions/articles/6000060969-how-is-the-altmetric-attention-score-calculated.

Many publications that critically target bibliometrics focus either on the method as a whole or on flawed bibliometric indicators (e.g., the *h* index, Hirsch 2005). We believe that many critical points are justified (see e.g., Marewski and Bornmann 2019) and should be considered in the use of and research on bibliometrics, but the points are not processed in coordinated research activities targeting what might be a valid reliance on bibliometrics in research evaluation. Bibliometrics lack a common, foundational research program which could function as an aegis for these activities. We think that an important condition for a common program is the existence of a conceptual framework (underlying the evaluative use of bibliometrics) which can be targeted by bibliometric research. The objective of this paper is to introduce the fast-and-frugal heuristics framework as a candidate conceptual lens on bibliometrics, so that the usage of bibliometrics in research evaluation can be conceptually understood, be empirically studied, and be effectively taught. Specifically, if the evaluative use of field-normalized citation impact indicator (e.g., the field-weighted citation impact indicator proposed by Elsevier in Scopus, see Elsevier 2016) and other bibliometric indicators are conceived of as heuristics, six interrelated areas of research emerge:

First, a conceptualization of these indicators as heuristics asks in what evaluation environment which indicator will yield smart judgments, and in which ones not. In contrast, the bibliometric literature has, thus far, focused on flawed bibliometric indicators which, according to this literature, should be fully abandoned or be replaced by more suitable indicators. A heuristic view on indicators would instead be *ecological*: just like any heuristic (see e.g., Gigerenzer and Brighton 2009; Mousavi and Gigerenzer 2017), per se hardly any indicator is "good or bad" (Gigerenzer and Gaissmaier 2011, p. 474); instead the performance of each indicator hinges on the task environment (see Waltman and van Eck 2016). Precisely describing the environments in which each heuristic works, in turns, is one of the key research goals of the fast-and-frugal heuristics framework (e.g., Gigerenzer et al. 1999; Todd et al. 2012).[5]

---

[5] Note that this ecological focus is one of the several lines of division between the fast-and-frugal heuristics and alternative approaches to studying decision making. The literature on *heuristics-and-biases* (e.g., Kahnemann et al. 1982), for instance, is often taken to suggest that heuristics are error-prone and biased mental shortcuts that rational decision makers ought to avoid (see e.g., Marewski et al. 2010a, b, for a discussion; see also Lopes 1991, 1992).

Second, the fast-and-frugal heuristics framework prescribes not only studying when which heuristic will 'work well' and when each will 'fail'; this framework also asks the question when and how people will use which heuristic (e.g., Bröder 2011; Gigerenzer 2008a; Gigerenzer et al. 1999, 2011; Marewski and Schooler 2011; Rieskamp and Otto 2006). Hence, *descriptive* work on bibliometrics would investigate when and how scientists, administrators, and others rely on field-normalized citation impact indicators and other indicators in decision making. Models of heuristics aspire to capture decisional (e.g., cognitive) *processes* (e.g., computational steps) and not only the *outcomes* (i.e., results) of problem solving (e.g., Hafenbrädl et al. 2016; for examples see e.g. Brandstätter et al. 2006; Bröder and Gaissmaier 2007). Those models (1) are cast in terms of algorithmic rules describing decisional processes; (2) ideally they come with a description of the skills and faculties (including cognitive abilities such as information-processing capacity) required to execute those algorithmic rules; and (3) ideally they specify precisely the (e.g., statistical) nature of problems which can be tackled (i.e., the task environments in which the rules can be applied; e.g., Gigerenzer 2008b; Gigerenzer et al. 1999; Gigerenzer and Gaissmaier 2011; Marewski et al. 2010a, d).

Certain construction principles of heuristics apply regardless of whether a heuristic represents a model of decision making an agent could be assumed to spontaneously adopt or a decision aid, engineered to help people make better decisions, with those people being trained or instructed to adopt the decision aid. For example, like many other heuristics, *fast-and-frugal decision trees* (short: *fast-and-frugal trees*) such as the one shown in Fig. 1 can be cast in terms of three sets of algorithmic rules (e.g., Martignon et al. 2008): A *search rule* that specifies what information (e.g., predictor variables) is searched for and how (i.e., in what order), a *stopping rule* that delineates when information search comes to an end, and a *decision rule* that determines how the acquired information is employed (e.g., combined) to classify *objects* (e.g., patients). When expressed more generally, the rules making up fast-and-frugal trees can be represented as follows (e.g., Gigerenzer 2007; Marewski and Gigerenzer 2012; for details see Martignon et al. 2008 or Phillips et al. 2017):

> *Search rule*: Consult predictor variables in a specific order (e.g., as a function of their positive predictive value and/or their negative predictive value).
> *Stopping rule*: Terminate information search immideately once one predictor variable is hit that permits classifying the object under consideration.
> *Decision rule*: Classify that object by relying on this predictor variable.

Skills necessary for using the fast-and-frugal tree shown in Fig. 1 include expertise in medical diagnosis, such as knowing how to assess changes in the ST-segment. The clinic environment in which this tree will yield accurate decisions must be specified in terms of the patient population and its attributes, because the same *classifier* (e.g., a fast-and-frugal tree, an HIV test, or a single bibliometric indicator used to categorize scientists) might yield different classifications, for instance, depending on the prevalence of a condition of interest within a population.

Third, several studies on fast-and-frugal heuristics have shown that their predictive accuracy can be similar to or higher than that of weighted-additive and other more information-greedy models (for overviews, see Goldstein and Gigerenzer 2009; Hafenbrädl et al. 2016; Marewski et al. 2010a). Let us mention just a few heuristics. Simply weighting information equally (e.g., Dawes 1979; Dawes and Corrigan 1974)—which is what the so-called *tallying heuristic* (see e.g., Gigerenzer and Goldstein 1996) does—can yield nearly as accurate inferences as multiple regression which 'optimally' weighs information; and in certain situations equal weighting might even end up representing the better option (e.g.,

Czerlinski et al. 1999; see also e.g., Einhorn and Hogarth 1975). The *take-the-best heuristic* (Gigerenzer and Goldstein 1996), which similarly to fast-and-frugal trees consults evidence (predictor variables) sequentially, has been found to outperform, on average, multiple regression across 20 different environments, constituted by data sets from psychology, sociology, demography, economics, health, transportation, biology, and environmental science (e.g., Brighton 2006; Czerlinski et al. 1999; Gigerenzer and Brighton 2009). The *recognition heuristic* (e.g., Goldstein and Gigerenzer 2002), which relies on name recognition as the only predictor variable to make inferences and forecasts, has been found to be able to aid decision making in geographical, sports, and other domains (e.g., Gigerenzer et al. 1999). For instance, a variant of that heuristic can predict the outcomes of Wimbledon tennis matches as well as or better than ATP rankings and seedings from Wimbledon experts (e.g., Scheibehenne and Bröder 2007; Serwe and Frings 2006). By exploiting people's systematic ignorance (i.e., their systematic lack of recognition of objects' names), a variant of that heuristic can, moreover, even aid to predict the outcomes of political elections (Gaissmaier and Marewski 2011).

Importantly, as can be seen from the examples enlisted above, prescriptive research on fast-and-frugal heuristics does *not* just investigate the absolute performance of a given heuristic, but asks how well a given heuristic performs in comparison to competing approaches (Gigerenzer and Brighton 2009; Gigerenzer and Gaissmaier 2011; Marewski et al. 2010d; Mousavi and Gigerenzer 2014)—including more complex algorithms, expert judgments, and other heuristics—in a given environment. That is, heuristics are typically not studied in solitariness. Prescriptive research on BBHs could follow a similar ecological benchmarking approach, examining in what environments and to what extent different research evaluation tools (e.g., peer review and alternative BBHs) perform well and when they do not.

Fourth, an adaptation of the concept of heuristics to evaluative bibliometrics seems reasonable, because bibliometrics are commonly used to assess a highly complex phenomenon, and one that comes with considerable uncertainty (see Marewski and Bornmann 2019). When will potentially different aspects of the scientific merit of an article, the achievements of a research unit or of a researcher reveal themselves: in a year from now, in five years, or perhaps never? How will those aspects come to light: will a finding lead to a revolution in technology and society (e.g., like the invention of steam engines or the computer), or will it 'simply' lead to new insights (e.g., about the workings of human memory)? By definition the future is knowable only from hindsight. But even the present and past can be uncertain, namely when the decision maker does not know the 'true state' of the world and has to *infer* that state.[6]

Heuristics are tools for dealing with uncertainty (see e.g., Gigerenzer 2008a; Gigerenzer and Marewski 2015). Yet, surprisingly, in the bibliometrics literature, only a few hints at heuristics can be found and those are not based on bibliometric-based, fast-and-frugal heuristics as potentially useful tools.

---

[6] In this paper, we use the terms "infer" or "inference" to refer to all judgments where the 'true state' of the world is unknown to the decision maker at the movement of making the inference—be those inferences about the present or past, or inferences about the future (=forecasts). That is, forecasts and predictions are also inferences (see also Gigerenzer and Gaissmaier 2011 for the fast-and-frugal heuristics research program's focus on inferences as opposed to preferences). Classifying research into different categories (e.g., 'high quality' versus 'medium' versus 'low quality') can warrant inferences, too (e.g., about quality; Marewski and Bornmann 2019).

For example, Saad (2006), Prathap (2014), and Moreira et al. (2015) conceive of the *h* index (and its variants, see Bornmann et al. 2011b) as a heuristic tool which reduces quantitative and qualitative information to a single value. Heinze (2012, 2013) introduces a heuristic tool (actually a classification scheme) "that singles out creative research accomplishments from other contributions in science" (Heinze 2012, p. 583). We have, in different contexts, hinted at heuristics ourselves and discussed related notions, such as decision making under uncertainty and the historical roots and societal context of the mindless usage of bibliometric indicators and other seemingly 'objective' statistics (Bornmann 2015; Gigerenzer and Marewski 2015; Hoffrage and Marewski 2015; Marewski and Bornmann 2019).

Beyond those isolated examples, to the best of our knowledge, people's use of heuristics in the area of evaluative bibliometrics has not been researched, and even less so empirically. It is unclear whether heuristics are applied and—if so—in what form. If the use of heuristics could be identified, their frequency of use could be detected in given task environments, and empirical investigations about the validity of judgments based on them could be undertaken. Just as in studies on heuristics in other areas, it could become the goal of descriptive and prescriptive research on bibliometrics to formulate and evaluate highly precise *formal* (i.e., computationally or mathematically specified) models that could be submitted to fine-grained mathematical analyses, powerful computer simulations, and strong experimental tests (for an introduction to formal modeling work, see Marewski and Olsson 2009): "formal models of heuristics allow asking two questions: whether they can describe decisions, and whether they can prescribe how to make better decisions than, say, a complex statistical method" (Gigerenzer and Gaissmaier 2011, p. 459).

<u>Fifth</u>, John Q. Public, experts, and other humble human beings are not endowed with "omniscience" and "omnipotence" (e.g., Gigerenzer 2008b, p. 4; Marewski et al. 2010a). They do not have complete knowledge of the past; they cannot foretell everything that will happen in the future; and their information-processing capacities (e.g., memory storage, computational power) are finite. Defying the (classic) rational economic theories of his time, Nobel Laureate and polymath Herbert Simon coined the term *bounded rationality* to refer to the limits of human ability. Sometimes overlooked, Simon not only focused on the bounds of human rationality (see e.g., Gigerenzer and Goldstein 1996; Marewski et al. 2010a): he also stressed that human capacities are adapted to their environment. Following Simon's footsteps, the fast-and-frugal heuristics research program has picked up those lines of thoughts (e.g., Simon 1955, 1956, 1990). Heuristics are environmentally-fitting tools for managing and reducing complexity, for instance, by allowing decision makers to focus on a few relevant objects or attributes (e.g., predictor variables) this way simplifying decision tasks. In conceiving of the fast-and-frugal heuristics program as a conceptual lens for studying bibliometrics in research evaluation, we suggest that those heuristics might ease the complex process of assessing scientific quality and deciding on units (e.g., on scientists, research groups, or countries; Bornmann 2015).

Specifically, research on BBHs could ask the question (i) to what extent (i.e., compared to other procedures) effort-reductions and time-savings occur, as well as (ii) when that is the case—that is, in what environments. As we will discuss in more detail below, the rationality of relying on heuristics depends not only on the environment, but also on the goals of decision making (Marewski et al. 2010a; Marewski and Gigerenzer 2012). Hence, fast-and-frugal heuristics are not just systematically evaluated in terms of one single performance criterion (e.g., accuracy when making inferences), but in terms of those criteria that match the task environment *and* goals at hand (Gigerenzer et al. 1999). For example, in some situations reducing effort, saving time, and/or making accurate inferences might be primary goals; in others (e.g., in strategic interactions), being transparent, predictable, and/or fair vis-à-vis cooperation partners might be what a decision maker cares most about (see also Gigerenzer and Gaissmaier 2011; Hafenbrädl et al. 2016).

Sixth, ever since citations are used as measures of quality or research impact, scientists have tried to formulate theories of citations (Bornmann and Daniel 2008; Moed 2017). Those theories do not focus on the evaluative use of citations, but on the process of citing: why do authors (researchers) cite certain papers and not others? Two prominent citation theories have been introduced hitherto (see overviews in Cronin 1984; Davis 2009; Moed 2005; Nicolaisen 2007). The first is Merton's (1973) *normative citation theory*: publications are cited because they have cognitively influenced the author of the citing publication. Merton's theory provides the theoretical basis for using citations in research evaluation, since citations indicate recognition by peers and the allocation of achievements to publishing authors. More citations mean more recognition and more attributed achievements. Publishing researchers are generally motivated to cite other researchers, because they belief it is fair and just to give credit to publishing (researching) authors.

The normative citation theory has been heavily criticized since its introduction. The most important critical points refer to the fact that the theory does not explain all citation decisions (or no real decisions at all). Other factors besides cognitive influence and peer recognition play a significant role, so many studies indicate (see the overview in Tahamtan et al. 2016). One of the earliest studies in this respect has been conducted by Gilbert (1977) who interprets citations as tools for persuasion. Authors select those publications for citing that were published by reputable researchers in their fields. Thus, it is not the scientific content of a paper that leads to citation decisions, but the anticipated influence of the cited author on the reader. These accompanying citations are intended to confirm the claims of the citing author. The reputation of the cited author and other factors influencing citation decisions are mostly regarded to be consistent with the *social-constructivist theory of citing* (see the overview in Cronin 2005), which views citations as potential non-scientific (e.g., rhetorical) devices. Such different theories of citation behavior can be extended in three interrelated ways when viewing citation decisions through the theoretical lens of the fast-and-frugal heuristics framework.

(i) For one, heuristics may be candidate models to explain citation decisions: what heuristics might researchers use (if any) to decide whom to cite and in which environment do they use each heuristic? Since the fast-and-frugal heuristics framework assumes people to come equipped with a repertoire of heuristics (Gigerenzer et al. 1999) that framework may allow, in principle, to reconcile both the normative and the social constructivist theory of citation decisions. For instance, one could ask the question in what situations researchers rely on heuristics that operate on recognition and achievement (as in the normative citation theory) and when non-scientific (e.g., rhetorical) considerations (as in the social constructivist theory) are key. Moreover, the fast-and-frugal heuristics framework offers computational (and mathematical) models of heuristics that might be able to combine both theories in unifying models. For instance, equal weighting and sequential (lexicographic) heuristics can consider multiple reasons for making decisions in a single mechanism.

(ii) Whereas the normative and social-constructivist approaches focus on the citing behavior of authors, heuristics may also model the behavior of those decision makers who use bibliometrics for evaluative purposes. A corresponding research program could identify which BBHs scientists, managers, and administrators rely upon and their reasons for using them. Furthermore, one could examine to what extent different research evaluation environments are suitable for using different BBHs. This evaluative dimension is new in bibliometric theory building, which has thus far, mostly focused on descriptive aspects of citation decisions.

(iii) Rather than conceiving of citation decisions in terms of one 'global construct' (e.g., 'persuasion'), the fast-and-frugal heuristics framework would ask fine-grained questions about how different environmental structures lead to the execution of different heuristics for deciding whom and how to cite, with each heuristic being cast as a precise computational model of the underlying cognitive processes (e.g., comprised of algorithmic search, stopping, and decision rules; see above). Precisely modelling the interplay of decision processes and environmental structure is important for studying bibliometrics as a research evaluation tool, as this interplay will influence the 'values' that bibliometric indicators take (e.g., citation rates, *h* indices, journal impact factors) in given environments. Put differently, the 'micro-level' decisions of publishing and citing scientists shape the 'marco-level' environments in which bibliometrics are applied for research evaluation purposes. Understanding how individual behavior (e.g., those of citing authors) shapes up different 'macro-level' environments might aid to model when different indicators will work well and when they will fail from a prescriptive point of view.

## Decision making under uncertainty: theory and implications

In the introduction, we presented contemporary examples of complex evaluation methods (e.g., 'U-Multirank'). Yet, the idea that challenging judgment problems such as reserach evaluation, ideally, warrant complex, multi-perspective, well-thought through (e.g., analytic) solutions instead of 'simplistic' (e.g., one-dimensional) heuristic shortcuts is not new—it is a widespread conception of the world that comes in many facets, with certain roots going back to the past century, and with others spreading even farther back into history, including to the Enlightenment. At the time, characters such as Pierre Fermat and Blaise Pascal, and later the Bernoulli cousins, Daniel and Nicholas, forged important elements of what are, nowadays, classic optimization (maximization) approaches to understanding 'rational' judgment and decision making (Brandstätter et al. 2006; Gigerenzer 2008b; Gigerenzer et al. 1999; Hafenbrädl et al. 2016; Hoffrage and Marewski 2015). As is frequently pointed out (e.g., Hafenbrädl et al. 2016), well-known contemporary examples of optimization come with the *maximization of* (*subjective*) *expected utility* and models of *Bayesian inference* (e.g., Arrow 1966; Edwards 1954; Savage 1954; von Neumann and Morgenstern 1947). One lemma of many of those classics and their derivatives is that rational (= optimal) decision making warrants full information; another is that 'best' problem-solving approaches can be identified—that is, approaches maximizing certain values (e.g., accuracy, revenue) while minimizing what would be 'mistakes' (see Gigerenzer 2008b; Hafenbrädl et al. 2016). For instance, in the extreme case, classic decision analysis suggests people (e.g., investors) can and should (a) consider all options (e.g., investments, courses of action) at hand, (b) enlist every single potential consequence that comes with deciding for an option (e.g., win 200, lose 250), and (c) assess all consequences' probabilities of happening (e.g., 50%, 97%). Finalizing the analysis, they should then compute the expected (e.g., monetary) value of each option in order to pick the 'optimal' one (i.e., the one that maximizes the monetary gain). Also, the Heart Disease Predictive Instrument mentioned above can be conceived of as a representative of optimization; and indeed, many statistical methods (e.g., regressions) used in science aim, in integrating information, to estimate 'optimal' coefficients (e.g., beta weights that maximize fit; Hafenbrädl et al. 2016; Marewski and Gigerenzer 2012).

When applying optimization methods to judgment problems, decision makers actually make (implicit or explicit) assumptions about the environment in which they decide (e.g., Hafenbrädl et al. 2016). As we have discussed in another paper, this also holds true for research evaluation (Marewski and Bornmann 2019). In the literature on fast-and-frugal heuristics, a shorthand to refer to one class of those environments is the term *risk* (Knight 1921): According to Gigerenzer (e.g., 2014), in *worlds of (known) risk*, options, consequences, as well as their "probabilities are known or can be reliably estimated" (Hafenbrädl et al. 2016, p. 216). In those "well-defined and predictable" (Hafenbrädl et al. 2016, p. 216) situations, (some form of) optimizing can not only be viable, but also be reasonable. Perhaps the most intuitive examples of such environments are gambles of chance (e.g., playing dice, roulette, and lotteries see e.g., Gigerenzer 2014). In theory, so one might believe, in those task environments, options in such games (e.g., betting on any number from 1 to 6 with a dice), their consequences (e.g., win $200 with a 6, lose $41 with any number from 1 to 5), and probabilities (e.g., 1/6) can be specified and/or be estimated, say in terms of frequencies (Gigerenzer 2014; Hafenbrädl et al. 2016; Mousavi and Gigerenzer 2014, 2017).

Yet, in practice even in gambles surprises may strike (e.g., fraud in dicing): *Small worlds* (see Savage 1954) of risk—situations where all the options, consequences, and probabilities are knowable—might be conceived of as one end of the spectrum (Gigerenzer 2014; Gigerenzer and Marewski 2015). Those at the other extremity are called *large* (or *uncertain*) *worlds* (see Binmore 2007, 2009; Savage 1954) in the literature on fast-and-frugal heuristics.[7] Following Gigerenzer (e.g., 2014; Hafenbrädl et al. 2016; Mousavi and Gigerenzer 2017), in large, uncertain worlds, decision makers may not recognize and know all their options, with some options resting forever 'in the dark'. Decision makers may also not be aware of all potential consequences of the options at hand. Likewise, the probabilities that any of those consequences will come about may simply belong to the domain of the fully unknown and/or there may not be enough information (empirical observations) to sufficiently accurately estimate those probabilities. As Hafenbrädl et al. (2016) point out, "[i]n such situations, surprises can occur, leaving the premises of a rational (e.g., Bayesian) decision theory unfulfilled. Not only does uncertainty lead to optimization becoming unfeasible or inappropriate, but it also invalidates optimization as a gold standard to which other decision processes are compared" (p. 217) as benchmark.[8]

---

[7] Savage (1954) distinguished between small(er) and large(r) worlds, and Knight (e.g., 1921) similarly between risk and uncertainty (see Binmore 2007). In this article, we use those terms interchangeably, albeit following Gigerenzer's (e.g., 2014; Hafenbrädl et al. 2016; see also Gigerenzer and Marewski 2015) definitions (for a comparison, see Mousavi and Gigerenzer 2014). Also, please note that the notions of risk and uncertainty might, *generally*, be useful ones when reflecting about research evaluation (that is, useful independent of whether one studies BBHs or not). For example, we have used those very notions to question the meaningfulness of research evaluation exercises, as well as to discuss research evaluation in terms of larger historical and societal trends (Marewski and Bornmann 2019).

[8] Indeed, there is a large literature (and a good amount of debate) about the scope of classic rational models and their (e.g., statistical) relatives (see e.g., Mousavi and Gigerenzer 2017; Shanks and Lagnado 2000; Todd and Gigerenzer 2000). For instance, Gigerenzer and Marewski (2015) point out that some authors (e.g., Lindley 1983) do believe that Bayesian statistics can be applied to *all* kinds of uncertainty, including single (e.g., isolated) occurrences: to paraphrase the line of reasoning, after all, one can specify subjec-

The conceptual lens of the fast-and-frugal heuristics framework suggests that most real-world decision tasks come, in essence, as situations of uncertainty: large worlds that may (or may not) entail elements of risk, but that even if they include some known risks, still remain fundamentally uncertain (Gigerenzer 2014; Hafenbrädl et al. 2016; Mousavi and Gigerenzer 2014, 2017). In our view, this also holds true for research evaluation tasks (Marewski and Bornmann 2019). For instance, imagine it might be possible to reliably assess what budget will be available to fund experimental research at your university during the next two years; it might also be known which projects are in need of funding (= your known 'options'). However, what might not be known or knowable is the likelihood that any of those projects will lead to an outcome (= a 'consequence') that justifies investing in it in the first place (Marewski and Bornmann 2019). Moreover, while some outcomes that could justify funding might be imaginable in advance, others might not be (e.g., think of a project unexpectedly leading to new technological developments). Finally, it might not be known that (with a little bit of lobbying) the same budget could alternatively be used for funding fewer and more expensive longer-term projects (i.e., stretching out beyond the two-year horizon). That is, the notion of uncertainty also includes situations where some, but not all options, consequences, and probabilities are known (Mousavi and Gigerenzer 2014, 2017).

To illustrate the fast-and-frugal framework's conceptual lens on decision making under uncertainty in more detail, let us consider inferences about quality (e.g., of departments, scientists, manuscripts, grant applicants) in science as example. Elsewhere, we have pointed out that, conceptually speaking, such inferences can be thought of as classifications (Marewski and Bornmann 2019),[9] and as with all judgments under uncertainty, these classifications are likely never perfect: evaluators may adequately infer high quality work (e.g., sound theory, meaningful empirical results) to represent high quality research (*correct positives*); likewise, they may correctly recognize 'bad' work (e.g., flawed analyses, trivial results, weak theory) as such (*correct negatives*). However, evaluators may also be misled and consider 'bad' work to be high quality research (*false positives*) and they may fully fail to recognize high quality research (*false negatives*) (see Bornmann and Daniel 2010). Those four possible classification outcomes can be visualized in a 2 × 2 table (often called confusion matrix). As mentioned above, in developing heuristics for classification tasks, the fast-and-frugal heuristics framework would suggest benchmarking different classifiers (e.g., fast-and-frugal trees based on bibliometric statistics) in model comparisons (see e.g., Gigerenzer and Brighton 2009; Gigerenzer and Gaissmaier 2011; Marewski et al. 2010d). This way it is, in principle, possible to find out which classifier yields

---

Footnote 8 (continued)

tive probabilities that aliens will land on earth or that Michael Jackson visited the moon—there are "no limits" (Gigerenzer and Marewski, p. 431) to coming up with subjective priors and using them in Bayes' rule. As Gigerenzer and Marewski (2015) point out further, other authors have more reservations (see also e.g., Mousavi and Gigerenzer 2017): To Savage (1954), the main proponent of modern Bayesian decision theory, an unlimited application of that theory seemed "utterly ridiculous" (p. 16)—In his view, "[i]t is even utterly beyond our power to plan a picnic", and that "even when the world of states and the set of available acts to be envisaged are artificially reduced to the narrowest reasonable limits" (p. 16)—those who plan an action can simply not be aware of every possible consequence that might occur as a result of the action. Following Gigerenzer and Marewski (2015), probably the least debatable area of application of Bayes' rule are those situations where data are available to empirically and reliably estimate probabilities, such as it can be the case in medical diagnosis, informed by systematic epidemiological observations.

[9] The points below are framed in terms of the fast-and-frugal heuristics research program; yet those are general points about classification under uncertainty that apply to any type of modeling (see e.g., Marewski and Bornmann 2019 for the same, following points, framed more generally).

desirable classifications for a given task environment (see e.g., Phillips et al. 2017, for a practical example and software for fast-and-frugal trees). This way it is also possible to develop families of classifiers or variants of the same model. For example, the four outcomes (e.g., true positives), can serve to estimate positive and negative predictive values, which in turn, can inform the ordering of predictor variables when developing and testing different (e.g., fast-and-frugal) classification trees.

A series of questions about ecological uncertainty would be raised on the way to conducting corresponding model comparisons. Notably, when evaluating different classifiers (i.e., in inference, see Gigerenzer and Gaissmaier 2011) an external (outside) criterion variable is needed. That criterion variable forms part of the environment. In the medical world, such a variable can be the actual health status (absence vs. presence of a given disease); in industrial quality control it can be likely product quality (zero manufacturing errors vs. at least one error), in customer care it can be the clients' future behavior (likely to stay with company vs. likely to leave to the competition) and so on (see Marewski and Bornmann 2019).

Criterion variables are fraught with uncertainties in many real world environments. In medicine, this can be the case, for instance, when there is no way of knowing, for sure, what disease has actually caused a patient's symptoms, or when diagnostic categories are fuzzy or overlap each other (as in psychiatry). Uncertainty can also arise when the criterion can, statistically, not be perfectly predicted from any set of predictor variables. In many real world domains, the relation between predictor variables and a criterion is obscured by noise; neither the predictors nor the criterion might be stable (but e.g., fluctuating in unexpected ways), and there might be unknown variables that shape the criterion and what are taken to be known predictor variables. The more sources of surprising 'errors' there are, the less predictable the criterion will be (see also e.g., Gigerenzer and Brighton 2009). Different readers of this paper will likely associate different types of scientific research (e.g., in psychology, education, economics, medicine, or metrology) that tries to predict behavior, performance, events, statistics, and other things from different kinds of (e.g., questionnaire, laboratory, field) data with different degrees of this kind of predictive uncertainty (e.g., predicting market or crime rate fluctuations from economic indicators, modeling epidemic deaths from health statistics). Also in research evaluation, criterion variables can come with considerable uncertainty (Marewski and Bornmann 2019): Leaving trivial cases aside, how frequently does one know *for sure* how good a piece of research really is? And if ever, *when* does one know this—a few months or years later, or after decades when the revolutionary ideas expressed in a paper finally are ripe to become recognized (Gigerenzer and Marewski 2015)? Such uncertainties are hinted at by low agreements between reviewers in journal peer review procedures assessing the same manuscript (Bornmann et al. 2011a).

Moreover, in research evaluation, the consequences (e.g., costs and benefits) associated, at $T_2$, with correct positive, correct negative, false negative, and false positive judgments might be difficult to assess at $T_1$, or even completely unknowable. As we have argued elsewhere (Marewski and Bornmann 2019), one can only speculate about the 'costs' (e.g., for society, a national funding agency, an institute, a researcher, patients) if a single landmark piece of research is classified as 'bad', or thousands of trivial, inconsequential empirical findings or even dubious or potentially dangerous (e.g., medical, drug-related) ones as excellent, high-quality work. Sometimes certain costs might be known (e.g., the amount of money invested in a project), but then the benefits coming from that investment might be hard to estimate (e.g., how to estimate the benefits of knowledge transfer to future PhD students?). Also in areas other than research evaluation, the consequences of decisions can often only be roughly understood: For example, when it comes to breast cancer screening (another type of classification), one can ask what costs and benefits are there from a health perspective (see Fig. 2); however such costs and benefits only become sufficiently predictable after large amounts of epidemiological data have been collected (Gigerenzer 2014). Other hidden costs might never be known. And if known, different costs might be difficult to be traded off against each other (e.g., money versus health) and/or come with new uncertainties (see Marewski and Bornmann 2019 for a more detailed discussion).

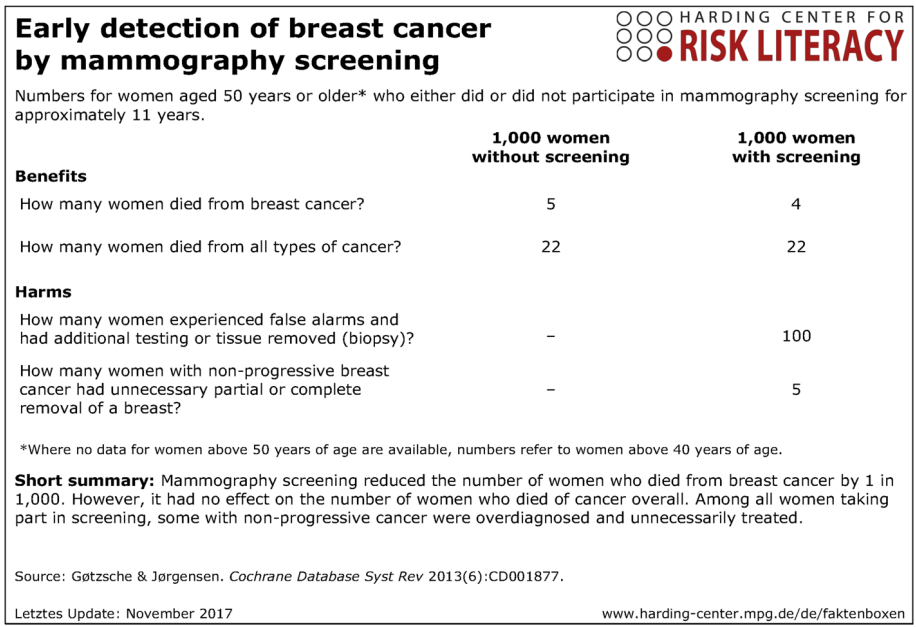

**Fig. 2** Benefits and harms (= 'costs') of mammography screening in comparison to no mammography screening. *Source*: https://www.harding-center.mpg.de/en/fact-boxes/early-detection-of-cancer/breast-cancer-early-detection

The fast-and-frugal heuristics framework suggests to *not* mistake unpredictable uncertainties for risks (e.g., Gigerenzer 2014; Gigerenzer and Marewski 2015; Hafenbrädl et al. 2016; Mousavi and Gigerenzer 2014, 2017). Instead, the framework takes uncertainties seriously, and in so doing, makes their existence transparent—potentially also in research evaluation (see Marewski and Bornmann 2019). The framework does also not pretend that optimization is *the* 'optimal' approach for managing uncertainty (e.g., Hafenbrädl et al. 2016). Moreover, the framework acknowledges that there might be very different answers to the same questions (e.g., about scientific quality), depending on what tools (e.g., different classifiers such as different BBHs) one selects (i.e., from the toolbox; see Gigerenzer et al. 1999, and below) and depending on what criteria and cost–benefit structures form part of the decision environment (e.g., Luan et al. 2011; Phillips et al. 2017)—again, potentially also in research evaluation (see Marewski and Bornmann 2019). What is more, in assuming a repertoire of decision strategies to act as tools for dealing with uncertainty, the framework allows for flexible solutions to decision problems. To illustrate that last point, when classifying scientific output (e.g., as 'high quality'), the actual goal might not be to make 'optimal' classifications, but to simply allocate funds to different projects and/or scientists, with judgments about 'scientific quality' being used to justify such funding decisions. Hence, instead of trying (or merely pretending) to infer quality, decision makers might directly resort to other heuristics for selection. That is, a problem of inference and classification is replaced by a selection problem.

What heuristics for selection exist? *Selection heuristics* can implement *satisficing* decision processes: the notion of satisficing, coined by Simon (1956, 1990), suggests that decision processes are not geared towards optimization (maximization) but towards 'sufficiency' in task environments (see e.g., Gigerenzer and Brighton 2009; Gigerenzer and Goldstein 1996). For instance, when it comes to searching and selecting a mate, rather than considering all possible sexual partners, a satisficing strategy would first set an *aspiration level* (e.g., based on past sequential encounters with different individuals), and then lead a decision maker to pick the first mate that meets that aspiration level (see e.g., Todd and Miller 1999, for a discussion). In science, areas of application for satisficing might be hiring (e.g., selecting a suitable candidate for a professorship), funding (e.g., identifying grant proposals worth being funded) or literature search (e.g., finding a citable paper). To elaborate on just one example: instead of trying to rank all scientific output (e.g., of units or scientists), and then trying to identify the 'costs' of false assignments of ranks, and so on, a funding body could simply define criteria (i.e., aspiration levels) for funding eligibility and then allocate money to all applications that meet those criteria. Also fast-and-frugal trees can be designed as selection heuristics and they have been investigated as descriptive models for personnel decisions (Luan and Reb 2017).

Heuristics for selection, and inference and classification can be complemented by *social heuristics*. Some of those heuristics may be rooted in our evolutionary history. Social heuristics such as *Do what the majority does!*, *When confronting a task, imitate those who are sucessful at it!* (see e.g., Gigerenzer 2008b, 2010; Gigerenzer and Brighton 2009; Hertwig and Hoffrage 2013) can be tools for situations when little is known about the task environment at hand, such as when the criterion variable is not directly accessible and/or reliable non-social environmental signals (predictor variables) are sparse or otherwise out of reach. As such, social heuristics may aid dealing with situations where individual feedback learning is not possible, difficult, or dangerous (see also e.g., Gigerenzer 2010; Gigerenzer and Gaissmaier 2011; for an entire volume on heuristics in a social world, see Hertwig et al. 2013). Danger can arise when immediate feedback would imply death (e.g., it is dangerous to try learning how to infer, by oneself through a reality-check, which mushrooms can be correctly classified as eatable), or when the criterion only reveals itself after a time lag (e.g., as in inferences about potential adverse health effects produced by long term exposure to near-ultraviolet blue light emitted by computer screens; see also Gigerenzer 2008b). Also in research evaluation, feedback may emerge only years later, for example, when revolutionary ideas become, slowly, appreciated within the scientific community (Gigerenzer and Marewski 2015).

## Four points key to the study of fast-and-frugal heuristics

Indeed, when scientists and evaluators seem to use citation counts as *cue* (predictor variable) for scientific quality, they might, essentially be relying on social (e.g., imitation) heuristics (e.g., *'Cite the paper on a topic with the most citations!';* see also section "How the study of heuristics can inform the study of bibliometrics"). Often, that might be a good idea, just like it can be smart for cattle in a herd to imitate one another when one animal starts, all of a sudden to run (e.g., that animal might have detected a predator). Yet, just like a heard of stampeding cattle might run off a cliff, also in research social heuristics will not always yield clever decisions—for instance, wrong theses might be propagated for a long time. This example and the line of reasoning exposed in the previous section serve to illustrate four general points key to the study of fast-and-frugal heuristics (e.g., Gigerenzer and Gaissmaier 2011; Marewski et al. 2010a, d):

The first one is that no heuristic will always lead to clever (or 'best') judgments—no decision mechanism (be it e.g., complex or simple), results in good decisions in all situations (Gigerenzer and Brighton 2009; Marewski et al. 2010a, b). Moreover, often all that matters is not whether decisions are optimal but whether they are satisfactory enough (e.g., on average) relative to an agent's goals, in a given task environment. Translated to research evaluation, this means that no evaluation tool, be it different bibliometric indicators or peer-review, should be expected to always yield good judgments (Marewski and Bornmann 2019).

The second one concerns the notion of the *adaptive toolbox* (Gigerenzer et al. 1999): Rather than assuming decision makers' cognitive mechanisms can be best described in terms of just one 'all-purpose', domain-general tool (e.g., one type of strategy) serving to solve every single problem they encounter (see e.g., Marewski and Link 2014), the fast-and-frugal heuristics framework posits that people (and animals) can adaptively select from a large repertoire of different, domain-specific strategies (= including both various heuristics and other more complex methods): each strategy (= each 'tool' in the 'toolbox') is tuned to, and hence suitable for, a given domain, that is, a certain task environment (see e.g., Gigerenzer 2008a; Gigerenzer et al. 1999; Gigerenzer and Brighton 2009; Gigerenzer and Gaissmaier 2011; Marewski et al. 2010a, d; Todd and Gigerenzer 2000).[10] Being able to smartly choose among the different tools from this toolbox as a function of the task environment at hand makes up the expertise (and art) of clever professional decision making (see e.g., Hafenbrädl et al. 2016). For instance, bibliometricians ought to know when to rely on which indicator, and when to not rely on bibliometrics at all and to switch to extensive peer-review procedures instead (Marewski and Bornmann 2019).

The third point concerns normative criteria for evaluating decisions: How should decisions be made? In many areas, conventional wisdom and/or classic models of decision making and their offsprings are taken to suggest that the more information integration, the better for finding optimal—and hence rational—solutions (see e.g., Gigerenzer and Brighton 2009). In this view, rationality can enter the scene with optimality. Also *coherence*, as embodied by the rules of logic, is a norm for rational decision making.

---

[10] The notion of an adaptive toolbox represents a framework, comprised of separate formal (e.g., computational) models (e.g., of different heuristics); but the toolbox itself is not a formal (computational or mathematical) theory. In order to qualify as such a theory, in our view, a series of requirements would have to be fulfilled. For instance, the notion of a toolbox would have to be formally integrated with models of how different strategies (e.g., heuristics) are selected as a function of the environment as well as with formal models of how basic cognitive capacities such as perception and memory represent the environment. That is, the notion would have to be implemented into a cognitive architecture (e.g., Anderson et al. 2004). This is a nontrivial modelling challenge, and given its magnitude, thus far, only very first (limited) steps have been undertaken to tackle it (e.g., Marewski and Schooler 2011). However, corresponding calls for more modelling work have, repeatedly, been formulated (e.g., Tomlinson et al. 2011). All that said, there may be no unanimity concerning the ways how the notion of an adaptive toolbox is used in the experimental research practice (Marewski et al. 2018). For instance, Scheibehenne et al. (2013) point out that a toolbox model "can be conceptualized as a set of different psychological processes or strategies $f$ and the different parameters $\theta_f$ that may be associated with each of them" (p. 41). Moreover, there are models of heuristic selection that can be experimentally tested (see Jekel and Glöckner 2018a; Rieskamp 2018, for a recent discussion).

In contrast, in assuming an ecological view of rationality (dubbed *ecological rationality*; e.g., Goldstein and Gigerenzer 2002; Todd et al. 2012), the fast-and-frugal heuristics framework places emphasis on *correspondence* with the environment (for the distinction between coherence and correspondence, see Hammond 1996): what matters is not whether a decision is in line with the potentially context-blind prescriptions of, say, logic, but to what extent that decision can aid an agent to solve a problem in a given (real world) task environment (e.g., Gigerenzer et al. 1999; Marewski et al. 2010a). Moreover, ecologically-rational solutions do not necessarily have to be perfect, but satisfactory for achieving a given goal.[11] And sometimes, an ecologically rational heuristic for one decision maker's task is a fully unsatisfactory or even harmful heuristic for another. Hence, no strategy is universally rational for all people and in all situations; instead the rationality of using a given strategy always has to be examined relative to a *specific person's goal* in a given task environment (see e.g., Gigerenzer and Brighton 2009; Gigerenzer and Gaissmaier 2011; Marewski et al. 2010a, d).

To illustrate the latter point, if physicians work in an environment where they risk being sued for oversights and mistakes, then it is ecologically rational for them to perform diagnostic tests and examinations on patients even if they think that those are unwarranted. This 'conservative' heuristic might not be beneficial for patients, because it increases the false-positive rate, potentially leading to harmful over-diagnosis and over-treatment. Yet, the heuristic protects the doctor from legal actions directed against her. We alluded to this type of defensive heuristic in the introduction, when we pointed out that a doctor might simply follow the rule of thumb to send all patients who exhibit serious chest pain to the coronary care unit (see Marewski and Gigerenzer 2012). 'Playing it safe' is an important goal in many professional environments (Artinger et al. 2019). In research evaluation, a prototypical heuristic might be to always assess other units (e.g., departments) positively when the review is non-anonymous. This defensive heuristic is a social one; using it helps to avoid making enemies. Or, in a public funding agency, an administrator may instruct the agency's evaluation panel what reasons for rejections of proposals need to be put in writing: the goal is not only to inform reviewers, but to make sure that enlisted reasons are legally bullet-proof so that applicants cannot sue the funding agency for rejections. In the humanities, another example comes in the form of evaluations based on bibliometric indicators (Ochsner et al. 2016). Those indicators might be used when one believes that numbers are socially more accepted than seemingly more subjective assessments (e.g., based on reading papers), even when knowing that the indicators might be less informative (e.g., than the contents of the papers read; see also Marewski and Bornmann 2019, for a discussion of obsessions with accountability in science).[12]

---

[11] In many domains, 'perfect' solutions may not be attainable in the first place. In inference, this is the case when criterion variables are highly unpredictable for any model, be it simple or complex. Then performing just a bit better than chance might be the best a decision maker can do. For instance, when inferring which of two options scores a larger value on a given criterion, the accuracy of inferences that can be made with the *fluency heuristic* (Schooler and Hertwig 2005) is often only slightly above chance level (e.g., 55%) in situations where decision makers have no additional knowledge about the options at hand (Marewski and Schooler 2011). However, since decision makers have no access to other information, they can do little more other than betting on the fluency heuristic (i.e., inferring more fluently retrieved objects to be larger) or making random guesses—the latter would lead to chance-level performance (50%). Thus, even though by absolute standards the heuristic's accuracy does not look impressive, when taking into account the decision situation, it becomes clear why relying on that heuristic can be ecological rational. Research evaluation might come with similar situations where achieving even just a minimum of performance can be ecologically rational.

The fourth point, key to the study of fast-and-frugal heuristics, concerns the research questions to be asked. Specifically, the study of heuristics asks *descriptive*, *ecological*, *applied*, and *methodological* ones (see Gigerenzer 2008a; Gigerenzer et al. 1999, 2008, 2011; Gigerenzer and Gaissmaier 2011; Marewski et al. 2010d; Mousavi and Gigerenzer 2014), all of which are relevant for the discussion of heuristics in the context of bibliometrics-based research evaluations.

(i) *Descriptive*: What are the contents and mechanisms of the adaptive toolbox: what heuristics do people use and when do they rely on which heuristic (e.g., Gigerenzer et al. 1999; Marewski et al. 2010a)? For instance, when will decision makers rely on imitation strategies, and when will they try to uncover solutions to decision problems themselves?[13]

(ii) *Ecological*: To which environmental structures are the different heuristics adapted, that is, when does utilizing a given heuristic lead to accurate, fast, effortless, defendable, or otherwise adaptive decisions *and when not* (e.g., Gigerenzer and Brighton 2009; Gigerenzer et al. 1999; Katsikopoulos et al. 2010; Marewski et al. 2010a; Marewski and Schooler 2011; Todd et al. 2012)? To illustrate this question, under what statistical conditions will a given fast-and-frugal tree lead to accurate classifications?

(iii) *Applied*: How can we help people to make better decisions (e.g., Gigerenzer et al. 2011)? If we know when people adopt which heuristic (descriptive question) and when different heuristics lead to accurate, fast, defendable, or otherwise adaptive decisions (ecological question), we can help people to make better decisions (e.g., Gigerenzer 2008a). For instance, decision making can be improved by changing (e.g., through policy-making) the environment so that the environment betters fits the heuristics used. Decision making can be improved, too, by changing (e.g., through training and instruction) the heuristics the decision maker uses so that the heuristics better fit the environment (e.g., Gigerenzer 2008a, b; Mousavi and Gigerenzer 2017). To give another example of the applied study of heuristics (see Hafenbrädl et al. 2016 for an overview), reconsider the kind of classification problems from medicine and research evaluation we described above—those where the criterion is not directly (or immediately) accessible. Rather than letting decision makers resort to selection (e.g., satisficing) or social heuristics, one can try to make the criterion more accessible (e.g., through measurement and data bases, such as by making the tools for bibliometrics-based institutional evaluations available; but see Marewski and Bornmann 2019 and Weingart 2005, for a critical discussion of the increasing quantification of science).

---

[12] In adopting such an ecological view on rationality, the study of fast-and-frugal heuristics can aid both: (i) uncovering such defensive social heuristics and (ii) modifying the corresponding decision making processes. To illustrate the first point, in modelling London judges' bail decisions with fast-and-frugal trees, Dhami (2003) was able to uncover that those judicial decisions were geared to enable the judges to "pass the buck" (see Gigerenzer and Gaissmaier 2011, p. 469; see also p. 467). As to the second point, Luan et al. (2011) offer a signal-detection analysis of fast-and-frugal trees that allows, as a function of the goals of the decision maker, to engineer either more defensive (conservative) or more liberal trees (see Hafenbrädl et al. 2016, for a discussion). When building bibliometric indicators into decision trees, the same kind of signal-detection analysis might help to create more conservative or more liberal classifiers, leading either to a larger probability of producing false negative or false positive research assessments (e.g., classifying, with a larger probability, 'high quality' work as 'bad' or 'low quality' work as 'good').

[13] Corresponding descriptive research on heuristics ideally not only tests models of heuristics, but additionally includes complex alternative approaches in model comparisons (Gigerenzer and Gaissmaier 2011). After all, the claim of the fast-and-frugal heuristics framework is not that simple decision mechanism will always be relied upon—rather, people adaptively switch between different decision mechanisms (including both complex and simple ones) as a function of the task environment. Descriptive research on heuristics hence also asks the question whether heuristics are good models of behavior in the first place (e.g., Glöckner et al. 2014). For instance, while the assumption that our cognitive make-up comes with a repertoire of mechanisms is common in many areas of psychology and biology (see Marewski and Link 2014, for an overview) the decision making literature has put forward alternatives to such *multi-mechanism* conceptions, too, including the proposal that one *single* mechanism is sufficient to explicate behavior across tasks (e.g., Busemeyer 1993, 2018; Glöckner and Betsch 2008). Which type of approach is theoretically more plausible is debated in the literature (see Marewski et al. 2018, for a special issue on the topic). To give another example, for years, the aforementioned recognition and take-the-best heuristics have been at the centre of critique, questioning both heir psychological plausibility and the evidence that people rely on those heuristics in decision making (e.g., Dougherty et al. 2008; Gigerenzer et al. 2008; Marewski et al. 2010c, e, 2011a, b).

One can also create environments in which making mistakes (e.g., using the 'wrong' heuristic) is likely less costly. Strategies for managing mistakes are well known in aviation and increasingly also in medicine as part of the safety culture (e.g., Helmreich and Merritt 2016). In research evaluation such strategies might bolster the consequences of misclassifications (e.g., evaluating 'good' work as 'bad')—in the extreme case one would eliminate all potentially negative ones (e.g., *Do not base tenure or funding decisions on bibliometric analyses!*; see Marewski and Bornmann 2019).

We hasten to add that the applied question also asks when optimization procedures should be relied upon: the toolbox of decision mechanisms consists not only of heuristics; also Bayesian models, subjective expected utility-maximization, and other complex models have a place in it and there are environments where complex approaches should be favored over heuristics. For example, Katsikopoulos (2011) developed a simple tree for deciding, more generally, when to rely on fast-and-frugal trees as opposed to more complex procedures, notably regression.

(iv)  *Methodological*: How can people's use of heuristics and the environments in which those heuristics work (and fail) be empirically studied (e.g., Gigerenzer and Brighton 2009; Gigerenzer and Gaissmaier 2011; Marewski et al. 2010d)? Research on heuristics makes use of a cannon of methods, including experiments and field studies (e.g., Bröder 2011), computer simulations (e.g., Czerlinski et al. 1999), and formal (e.g., mathematical) analysis (e.g., Luan et al. 2011; Martignon and Hoffrage 1999). The methods used are in line with the framework's theoretical tenets. For instance, experimental work in the lab should try to capture actual decision making environments in the real (outside) world, and not just focus on abstract, artificial (lab) tasks (Gigerenzer 2010; Marewski et al. 2010d). Moreover, different heuristics and more complex strategies need to be pitted against each other in model comparisons in different environments, and not be tested in solitariness (e.g., Gigerenzer and Brighton 2009; Gigerenzer and Gaissmaier 2011; Marewski and Mehlhorn 2011). An example of isolated 'tests' is the pitting of a model of a heuristic against a verbally-specified null hypothesis rather than against other formal models of decision processes (e.g., Mousavi and Gigerenzer 2017).[14]

## Reasons for using bibliometrics-based heuristics (BBHs) instead of other strategies for research evaluation

One might wonder why bibliometrics have occupied such a prominent role in research evaluation. Other indicators such as the amount of raised third-party funds or number of academic awards or presentations could have become equally popular—yet they did not.[15] Similarly, one may also ask why BBHs ought to be studied; strategies for research evaluations built from other indicators might also do the job. However, in practice, bibliometrics have certain advantages compared to other indicators:

(i) The data are available in large-scale databases (e.g., WoS or Scopus covering worldwide publications and citations in all fields).
(ii) The data can be used for empirical analyses either within single fields or for cross-field comparisons.
(iii) Bibliometrics are rooted in the research process of nearly every researcher. It is the task of researchers to make their results publicly available (this differentiates pub-

---

[14] In social science research, it is common to evaluate how well different models explain existing data and given observations. A typical case is fitting a regression model to data at hand, and reporting $R^2$ or some other measure to establish how good the model is. In contrast, the performance of fast-and-frugal heuristics for inference and classification is evaluated in *foresight*, that is, out of sample or out of population, reflecting actual decision making under uncertainty about the future or unknown (e.g., Gigerenzer and Brighton 2009; Gigerenzer and Gaissmaier 2011). It is in such predictions where heuristics can outperform more complex models (e.g., Czerlinski et al. 1999; Gigerenzer and Brighton 2009; for further explanation of the difference between fitting and predicting, see e.g., Gigerenzer 2004; Marewski et al. 2010a; Marewski and Olsson 2009; Mousavi and Gigerenzer 2014; Pitt et al. 2002).

[15] Yet, a completely different class of indicators might focus on methodological aspects of research quality. To illustrate this, one could count how many of the papers produced by a unit offer sufficiently complete descriptions of methods (e.g., data collection, data analysis) in order to allow for replication studies to be conducted, or how many papers avoid improper statistical rituals (e.g., "null ritual" in null hypothesis significance testing, Gigerenzer 2004, p. 587; see also e.g., Gigerenzer and Marewski 2015), or how many papers correctly distinguish between mere data fitting and actual predictions (e.g., with fixed parameter values; see Pitt et al. 2002). In fields such as medicine, the amount of pre-registered studies could also be taken into account. Corresponding indicators could be built into heuristics and other strategies for research evaluation, and be systematically studied through the lens of the fast-and-frugal research program. One such heuristic might be the *f*-index, suggested by Gigerenzer and Marewski (2015). As they point out, the *f*-index can serve to gauge to what extent a paper contains exercises of hypothesis *finding*, but problematically sells that hypothesis finding as if it were hypothesis *testing*. Corresponding 'scientific fishing expeditions' (hence the '*f*' in *f*-index), so the argument goes, take place when authors explore their data to find 'significant' results (e.g. $p < 0.05$), but frame those 'findings' in a way that the 'findings' rather seem to be the fruits of tests of actual hypotheses, specified prior to analyzing the data (Gigerenzer and Marewski 2015; see also Gigerenzer 2004; Kerr 1998). Search rule: *Count the number of hypotheses stated in a paper and the number of statistical tests, including p-values and confidence intervals computed*. Decision rule: *If more statistical tests are computed than hypotheses stated, then there is a cue suggesting that hypothesis finding may be disguised as hypothesis testing*. A variant of the *f*-index might also serve to evaluate the 'quality' of scientific journals, namely by comparing the number of statistical tests to the number of hypotheses stated in articles from a given journal.

(iv)  lic research from industrial research) and to embed the results in previous results (research stands on the "shoulders of giants", see Merton 1965).
(iv)  Bibliometrics substantially correlates with other indicators (see e.g., Gralka et al. 2019), including with those that are usually not available in large-scale databases (see point *i* above) or that more indirectly reflect research processes (see point *iii* above).
(v)  If simple indicators such as the number of publications or citations are used in research evaluations, researchers are able to interpret the results based on their own (long-standing) experiences in the field. Since researchers are usually publishing researchers, they pick up publication practices and citation cultures in their careers. They know which journals are reputable, and they have an intuition about which number of publications and citations is high or low in their field.
(vi)  Since bibliometric data are available for many researchers and science managers, they can be used when time constraints exclude complex methods for research evaluation, when researchers have to assess research activities outside of their own area of expertise, or when they evaluate many and big units (e.g., countries). In other words, bibliometrics can be applied to situations fraught by capacity limitations. In these situations, bibliometrics might provide a 'ballpark figure' of research performance. Using a bibliometric tool might be "ecologically rational to the degree that it is adapted to the structure of …[the] environment" (Gigerenzer et al. 1999, p. 13), and fitting the decision maker's goals.
(vii)  Bearing, perhaps, a certain resemblance to accuracy-effort trade-offs often assumed in other areas of decision making (see Gigerenzer and Brighton 2009; Gigerenzer and Gaissmaier 2011 for a discussion), the availability and use of bibliometrics can even be thought of in terms of a 'rational' trade-off. Not every research assessment is important enough to warrant spending a lot of time; thus, researchers choose bibliometric indicators, such as the field-normalized citation impact indicator, that save efforts. To obtain a rough impression of the research performance of researchers from a department, simple publication and citation numbers can be considered, albeit while taking the researchers' different academic ages into account (Bornmann and Marx 2014).

Indeed, there is empirical evidence that simple procedures, such as counting the number of publications or citations, lead to similar results as the more time-consuming and resource-intensive peer review procedure. Auspurg et al. (2015) investigated the research rating process by the *German Wissenschaftsrat* (German science council[16]) which compared and appraised the performance of German universities and non-university research institutions. The authors focused on the research rating in the field of sociology including more than 200 institutional research units. Between 2006 and 2008, 16 reviewers (well-known sociologists) evaluated these units in an informed peer review process; each reviewer had to perform up to 80 single judgments (on an ordinal scale: excellent, very good, good, and satisfying). The peer review process was very sophisticated; the group of reviewers met at 15 days. The workload per reviewer has been estimated with two to three months. In their empirical analysis, Auspurg et al. (2015) focused on the institutional assessment of research quality by peer review. They performed regression analyses for identifying how much variance in the judgements of the reviewers could be explained by bibliometric measures (and other indicators which have been controlled). The reviewers tried to abstain from sheer metrics-based judgements: Although they took institutional indicators (including basic bibliometric ones) into account, they made the unique effort to read selected publications that had been submitted by the evaluated institutions. Despite the reviewers' efforts, the results of the regression analyses by Auspurg et al. (2015) suggests that the research performance (quality) of the evaluated units—as determined by the group of reviewers—could be similarly measured by simpler compilations of bibliometric data (number of publications and citations). In other words, it seems that heuristics based on pure bibliometric data might have led to similar evaluations of the research performance of German institutions in sociology as more cumbersome peer review procedures—at the very least, those results motivate asking that question.

---

[16] See https://www.wissenschaftsrat.de.

A similar conclusion can be drawn from the empirical results of another study. Diekmann et al. (2012) examined how a jury awards prestigious prices to articles: citations point to exactly the same praiseworthy articles as the jurors do (alike results are published by Jansen 2012). Specifically, Diekmann et al. (2012) investigated "whether journal articles winning a prestigious award of the Thyssen foundation have a higher impact in the scientific literature than a control group of non-awarded articles. [O]n average, awarded articles harvest significantly more citations than articles in the control group. Most remarkably, the average citation rank exactly matches the rank order of awards. The top award earns most citations while the second award, the third award and the non-awarded articles exhibit citation counts in declining order" (pp. 563–564). In their study, the authors also considered that receiving a prestigious award could be *the reason* for higher citation counts rather than an indicator of scientific quality. Hence they focused the citation analysis on the period between the publication date and date of price awarding. Since the citation analyses exactly matched the results of the peer review process, reviewers (or the Thyssen foundation itself) might as well have relied upon citations as heuristic to select the price-winning articles.

Pride and Knoth (2018) investigated the relation between peer reviews and citation data at an institutional level using data from the *Research Excellence Framework* (*REF*) 2014. The REF is the "UK's system for assessing the excellence of research in higher education institutions" (see http://www.ref.ac.uk). The study is based on a dataset including 190,628 academic papers in 36 disciplines submitted to UK REF 2014 and institutional/discipline level peer-review judgments (154 institutions, 36 Units of Assessment, UoAs). The study found strong correlations between peer-review judgements and citation data. The results suggest—according to Pride and Knoth (2018)—that "citation-based indicators are sufficiently aligned with peer review results at the institutional level to be used to lessen the overall burden of peer review on national evaluation exercises leading to considerable cost savings" (p. 195). Traag and Waltman (2019) used the same dataset also to investigate the relation between peer review and bibliometrics at the institutional level. They found a relatively minor difference between metrics and peer review in clinical medicine, physics, and public health, health services and primary care. Similar results have been published by Harzing (2017).

Leaving potential caveats aside (see also footnote 14), the studies by Auspurg et al. (2015), Diekmann et al. (2012), Pride and Knoth (2018), Traag and Waltman (2019), and Harzing (2017) could be interpreted as seeming to exhibit—similarly as in other areas of research on heuristics—a 'less-is-equal effect': a complex procedure uses more information than a simple tool and performs extensive assessments, but nevertheless lead to similar final judgments. Precisely specifying when, how, and why such effects emerge could be the goal of a foundational research program on bibliometrics. Such a research program could also ask the more daring questions whether, when, how, and why less information can be even better: since such "less-is-more effects" have been documented in reserach on fast-and-frugal heuristics in other areas (Gigerenzer and Gaissmaier 2011, p. 453); could they also emerge in research evaluation if one develops and tests proper BBHs? That is, could it be that, under certain circumstances, less 'effort' leads to more adequate judgments?

## Bibliometrics-based heuristics (BBHs) illustrated: one-reason decision making in research evaluation

We propose to reflect about bibliometrics in terms of the fast-and-frugal heuristics research program in order to study it through the lens of a theoretical framework. Typically, bibliometric indicators are used as shortcuts in research evaluation—shortcuts that seemingly enable assessing research units in a short time with minimal effort. Furthermore, evaluative bibliometrics are frugal, since they hinge on minimal information in decision making and ignore the rest. Research achievements are reduced to publication and citation numbers without considering further indicators, such as number of grants, editorships, or contributions to conferences (for an overview of existing indicators, see Montada et al. 1999).

Gigerenzer and Gaissmaier (2011) proposed describing heuristics in terms of four broad classes of models: "The first class exploits recognition memory, the second relies on one good reason only (and ignores all other reasons), the third weights all cues [i.e., predictor variables] or alternatives [i.e., options] equally, and the fourth relies on social information" (p. 459). Largely drawing from their overview, in the following section, we discuss only a limited number of *one-reason decision making heuristics* for illustrative purposes. Other heuristics (one reason or not) that, in the context of evaluative bibliometrics, might (in the future) be interesting to study are, for example, profiling ones. Geographic profiling is, for instance, used "to predict where a serial criminal is most likely to live given the sites of the crimes "(Gigerenzer and Gaissmaier 2011, p. 463). Spatial bibliometrics (see an overview in Frenken et al. 2009) is an emerging topic in scientometrics with increasing popularity. Akin to profiling heuristics, spatial bibliometrics can be relied upon to identify hot and cold spots in international research (Bornmann and de Moya Anegón 2019).

What are one-reason heuristics? As mentioned above, social heuristics may be useful when little information and sparse (e.g., individual feedback) learning opportunities are available; the resulting uncertainties can let the 'turning to others' (e.g., to peers), say for observation and imitation, become a potentially sensible course of action (Gigerenzer and Gaissmaier 2011). In other situations, decision makers have sufficient knowledge to instead rely on (non-social) one-reason heuristics. A representative of one-reason decision making strategies is the aforementioned take-the-best heuristic (Gigerenzer and Goldstein 1996). Take-the-best is a model of inference. The heuristic prescribes sequentially considering predictor variables (called cues) in the order of their predictive accuracy (called *validity*), and bases a decision on the first cue that differentiates between options (e.g., Gigerenzer and Gaissmaier 2011):

> *Search rule*: When infering which of two options (e.g., two companies) scores higher on an uncertain criterion (e.g., future revenue), consider cues in the order of their validity, starting with the most valid (predictive) cue.

*Stopping rule*: Immediately once a cue is found on which one but not the other option has a positive cue value terminate search—without considering any other cues.
*Decision rule*: Decide for the option with the positive cue value as the one that likely scores a larger value on the criterion.

Leaving aside the possibility of using social information to derive cue orders for take-the-best (see Gigerenzer et al. 2008), this form of one-reason decision making hinges upon expertise, and hence information and/or learning opportunities: to use take-the-best, decision makers need to know which cues are the most predictive ones in order to sequentially consult them (see also Garcia-Retamero and Dhami 2009; Gigerenzer and Gaissmaier 2011; Gigerenzer et al. 2008).

Another sub-class of one-reason decision making heuristics seeks out just one 'clever' cue and bases decisions exclusively on that predictor variable without (e.g., sequentially) considering any others (Gigerenzer and Gaissmaier 2011). To illustrate this type of one-clever-cue heuristic, in business, Wübben and Wangenheim (2008) examined, empirically, different approaches for developing and implementing customer base management strategies. Such strategies are important to managers and other business professionals; they aid them to decide into which customers to invest "marketing budgets"—be it by offering certain customers discounts or by sending them, say, colorful catalogues (see Goldstein and Gigerenzer 2009, p. 766). Wübben and Wangenheim (2008) observed (i.e., in a small sample) that certain business professionals may rely on a "simple recency-of-last-purchase rule" (Gigerenzer and Gaissmaier 2011, p. 455), the *hiatus heuristic*, for customer classification.

*Decision rule*: "If a customer has not purchased within a certain number of months (the hiatus), the customer is classified as inactive; otherwise, the customer is classified as active." (Gigerenzer and Gaissmaier 2011, p. 455)

Note that in contrast to fast-and-frugal trees that operate on multiple predictors, the hiatus is the only cue considered in this classification strategy. To give another example of one-clever cue decision making, Wübben and Wagenheim (2008) also empirically investigated to what extent a different cue allows making a different bet, namely when it comes to forecasting who will be the "future high-value customers" (Goldstein and Gigerenzer 2009, p. 767).

*Decision rule:* Infer "that the top X% of customers in the past will continue to be the top X% of best customers in the future." (Goldstein and Gigerenzer 2009, p. 767)

In data sets from airline, apparel, and music (online CD retailing) business areas, this one-clever-cue heuristic has been found to single out best customers surprisingly well compared to two more complex models from the marketing literature (Wübben and Wangenheim 2008).

One-clever-cue heuristics seem common—not only in humans. As Gigerenzer and Gaissmaier (2011) note, "[m]any animal species appear to rely on a single 'clever' cue for locating food, nest sites, or mates. For instance, in order to pursue a prey or a mate, bats, birds, and fish do not compute trajectories in three-dimensional space, but simply maintain a constant optical angle between their target and themselves…" (p. 463). The (frequent) use of certain bibliometric indicators by researchers and science managers in decision making might be consistent with the definition of certain one-clever-cue heuristics, too.[17] For example, the field-normalized citation impact of papers is used to get a rough impression of a researcher's academic performance. The share of a university's papers which belong to the 10% most-frequently cited papers in their subject category and publication year is used in the Leiden Ranking to identify the 'best' universities worldwide. Moreover, the number of papers published by a research group is often used to estimate its 'productivity'.

In their own practice, readers of this paper will have come across many other such situations where one variable (one bibliometric indicator) is used in research evaluation. In our opinion, good reasons exist for studying to what extent (including: if at all) and when relying on the number of publications, citations, and other bibliometric indicators (e.g., as in one-clever-cue heuristics) may be ecologically rational. Research results that are not published cannot contribute to the archived knowledge in most disciplines. Thus, publishing results is an essential part of being a researcher (in most of the disciplines). A large number of publications from a researcher can signal that this researcher is active in many research projects, collaborates successfully with other researchers (co-authors), and received funds for different projects. In other words, the number of publications is related to other indicators, such as research funds received and (international) collaborations. Furthermore, it is essential for publishing scientists to embed their own results in the archived knowledge by citing the corresponding publications from other researchers. Citations can reveal the cognitive influences of the cited scientists and the roots on which citing scientists' research stands (see section "How the study of heuristics can inform the study of bibliometrics"). Not without reasons, do bibliometricians regard citations as a measure for one aspect of quality: the impact of research on research. Other aspects are accuracy and importance (Martin and Irvine 1983).

Since a few years, what one might call an (adaptive) toolbox for using bibliometric methods in research evaluation seems to be emerging. For instance, Todeschini and Baccini (2016) published the *Handbook of Bibliometric Indicators: Quantitative Tools for Studying and Evaluating Research*; Waltman (2016) reviews the literature on size-dependent and size-independent citation impact indicators. Most evaluative tasks focus on the output and citation impact of scientific units (e.g., of single researchers). When comparing different researchers from the same field, an output-oriented one-clever-cue heuristic might be:

*Search rule:* Search for all substantial publications (e.g., articles and reviews), produced by two researchers (A and B).
*Stopping rule:* Stop search once all such publications have been identified.
*Decision rule:* If scientist A has published more substantial publications than scientist B and both scientists are of the same academic age, then infer that scientist A is more active in research than scientist B.

This heuristic may work, so one could speculate, in academic environments where different authorship orders play no role (or 'average each other out') (Bornmann and Marx 2014).

It is standard in bibliometrics to measure impact in cross-field comparisons with field- and time-normalized citation scores. The *Leiden Manifesto* (Hicks et al. 2015) recommends the percentile indicators as the most robust normalization method: "Each paper is weighted on the basis of the percentile to which it belongs in the citation distribution of its field (the top 1%, 10% or 20%, for example)" (p. 430). The share of top $x$% papers for units in research can allow making a decision on their 'standing': higher values than $x$ might point to units with an 'above average' impact in the corresponding fields and publication years. Thus, the decision rule for an impact-oriented heuristic for institutional assessments might be:

> *Search rule:* Search for all substantial publications (e.g., articles and reviews), produced by the institute.
> *Stopping rule:* Stop search once all such publications have been identified.
> *Decision rule:* If the institute has published papers over several years with more than $x$% (e.g., say, 20%) top 10% papers in the corresponding fields and publication years, the institute has a (significantly) better performance than an 'average' institute in the world.

---

[17] Our use of the term 'one-clever-cue heuristic' might suggest that the application of bibliometric indicators is 'clever' in general. However, the only reason of using this label here is that the term has been proposed in the literature for a certain type of heuristic. Research on bibliometrics-based heuristics has yet to reveal whether and in which environments using different BBHs actually is 'clever'.

This decision rule somewhat resembles the hiatus heuristic mentioned above. Of course, using this heuristic would come with certain assumtions about the (e.g., distributional) structure of the world (and the meaningfulness of the notion of 'above average' in the context of research evaluation).[18]

Importantly, just as with other heuristics in the adaptive toolbox, the ecological rationaly of relying on any one-reason BBHs depends on environmental structure and the goals of the decision maker. When a decision maker wants to make predictions, outside the realm of research evaluation, statistical aspects of the environmental structure can include, for instance, "1. Uncertainty: how well a criterion can be predicted. 2. Redundancy: the correlation between cues [i.e., predictor variables]. 3. Sample size: number of observations (relative to number of cues). 4. Variability in weights: the distribution of the cue weights (e.g., skewed or uniform)" (Gigerenzer and Gaissmaier 2011, p. 457; for an entire book focusing on the ecological rationality of heuristics, see Todd et al. 2012). With respect to using bibliometric indicators to make predictions, one could ask, for example, to what extent (or not) one-reason heuristics might be useful in environments in which it is difficult to foresee research quality (e.g., in the evaluation of greater units, such as institutions or countries), and in which bibliometric indicators are highly correlated with other indicators of research performance (this is also especially the case with greater units). One could also ask to what extent (and under what conditions) the number of units that have to be assessed (e.g., many institutions or countries) might shape the usefulness of corresponding heuristics.

We hasten to add that, as any other method of decision making, naturally also one-reason BBHs will lead to errors even when they are relied upon in particularly fitting task-environments. However, it is an empirical question to what extent that error rate is larger than that of methods of judgment which follow other (e.g., more complex) rules.

Leaving such ecological considerations aside, from a descriptive point of view, it does not seem unreasonable to suspect that one-reason BBHs might be frequently used in research evaluation: these days, research must be evaluated under great time pressure.[19] Outside of the area of research evaluation, much (e.g., experimental) work has focused on the role of time pressure as one of many possible task-related determinants of decision making, fueling a multi-faceted literature (e.g., Bobadilla-Suarez and Love 2018; Glöckner and Hodges 2011; Marewski and Schooler 2011; Pachur and Hertwig 2006; Payne et al. 1988; Rieskamp and Hoffrage 2008). For research evaluation, one could start out by examining to what extent, and if so when and why, different BBHs may be more likely to be relied upon (or not) when the time available to make decisions varies from little to abundant. Furthermore, to what extent bibliometrics can, in the first place, realistically be replaced by other evaluation strategies may depend on (multiple) other aspects of the evaluative task environment at hand. If only a few research institutes with different missions (i.e., goals, see section "Four points key to the study of fast-and-frugal heuristics") must be evaluated, an informed peer review process including a thorough indicator report—reflecting the different missions—may be feasible to conduct. However, evaluators may actually be forced to resort to simpler methods when they face the challenging task of evaluating numerous institutions or scientists with the same mission: producing high-impact papers. When considering multiple options (e.g., numerous institutions), heuristics can be particularly helpful: early on—in a first step—in the decision process, heuristics can be relied upon to downsize the set of options (e.g., Gigerenzer and Gaissmaier 2011; Marewski et al. 2010a). Such *consideration-set* (e.g., Alba and Chattopadhyay 1985; Hauser and Wernerfelt 1990) generating decision processes have been described in consumer choice (see Gigerenzer and Gaissmaier 2011) and even in election forecasting (Marewski et al. 2010e). A classic model of choice that systematically eliminates options and that our readers might be familiar with has been developed in the 70ies by Amos Tversky (1972); it is tellingly called *elimination-by-aspects*.

---

[18] Note that with 'above average' not necessarily a statistical term is meant. This term is used often in research evaluation to refer to remarkable or non-standard performance.

Paralleling consumer and other choice situations, foundations for the promotion of research often receive more applications (e.g., for post-doctoral fellowship programs) than can be processed in a thorough peer review process (Bornmann and Daniel 2005). Thus, it is necessary to perform a pre-selection to reduce the number of candidates to be evaluated in more detail. The results of Horta and Santos (2016) suggest that "those who publish during their PhD have greater research production and productivity, and greater numbers of yearly citations and citations throughout their career compared to those who did not publish during their PhD" (p. 28). Similar results have been reported by Pinheiro et al. (2014) and Laurance et al. (2013). Based on the findings from such studies, a one-clever-cue heuristic for the pre-selection of applicants could prescribe selecting those who have published the most papers (articles or reviews) during their PhD:

> *Search rule:* Search for all substantial publications, produced by the candidates during their PhD.
> *Stopping rule:* Stop search once all such publications have been identified.
> *Decision rule*: Pre-select those candidates who have published most papers.

This heuristic might work in environments where applications focus on just one scientific field (e.g., decision making), implying that publication practices would be similar across applicants. Reversely, this heuristic might not be well-suited in task environments where, say, historians compete for grants alongside with neuroscientists: The latter will, likely, have more publications at the end of their PhD. An analogue ecological line of reasoning would apply if a similar heuristic were relied upon to select units worth additional funding (e.g., as in the *Excellent Initiative* in Germany[20]): choosing units worthy for consideration of additional funding based on the past numbers of substantial publications might be ecologically rational in environments where units have the same missions, but not where missions differ (e.g., publish basic research versus producing industry innovations, patents, or other 'applied output').

---

[19] That said, to the best of our knowledge, no large scale surveys have been conducted to test this assumption.

This pre-selection heuristic could by further refined—based on the results by Bornmann and Williams (2017) as well as Cole and Cole (1967)—by considering citations or journal metrics in a second step: Pre-selected candidates could be winnowed down further by identifying those with most publications in reputable journals or by zooming in onto those candidates with a minimum amount of citations.

Selection heuristics could also aid in the last decision round of a peer review process. In these last rounds, foundations face the problem of choosing among candidates all of whom are all well-suited, but available funds are insufficient for supporting all. Funders experiment with a lottery system to select among these candidates (Bishop 2018). The problem with lottery systems is that they do not rely on scientific quality criteria. BBHs could be an interesting alternative. For example, those candidates could be selected who have published the most highly-cited papers (i.e. papers belonging to the 10% most frequently cited papers within their field and publication year). This indicator is seen as the most robust field-normalized indicator (Hicks et al. 2015).

## Summary and Outlook: Towards the study of bibliometrics-based heuristics (BBHs)

The fast-and-frugal heuristics research program offers a theoretical framework for understanding how people can make ecologically rational decisions. Those decisions can be predictions about the future as well as inferences about the past or present (e.g., as in rankings, classifications, or estimations). Those decisions can also concern choice (e.g., of different courses of action), or search and selection (e.g., of candidates or mates), to name just two other examples. Many heuristics bet on a single cue (or few cues) among the many available cues for decision-making. Other heuristics draw on many predictor variables, but weigh them equally (e.g., as the tallying heuristic; see section "How the study of heuristics can inform the study of bibliometrics"). Those heuristics simplify tasks by ignoring order (ordering can be mathematically conceived of as a form of weighting; Gigerenzer and Brighton 2009; Gigerenzer and Gaissmaier 2011).

In contrast to the widespread view that complex tasks warrant complex, sophisticated solutions and that simple solutions reduce accuracy (and/or involve other trade-offs), the study of heuristics has shown—to speak with Gigerenzer and Brighton (2009)—that "less information, computation, and time can in fact improve accuracy" (p. 107). Decision makers—as a rule—often do not know all the relevant data (e.g., inherently limited information collections); they may have limited time and they have limited information-processing ability (e.g., working memories; see e.g., Simon 1990; see e.g., Marewski et al. 2010a, d for couching that point in terms of the fast-and-frugal heuristics framework). They frequently operate in environments with criteria (outcomes) that are difficult to predict (e.g., decision makers in science who try to select the next breakthrough research projects). In these uncertain environments, "less information can be more" (e.g., Marewski et al. 2010a, p. 111) when it comes to making inferences—that is, judgments about something that is unknown, be it because the criterion to be inferred lies in the future or because it is simply not accessible at present (see e.g., Gigerenzer and Brighton 2009; Gigerenzer and Gaissmaier 2011; Goldstein and Gigerenzer 2009; Hafenbrädl et al. 2016; Marewski et al. 2010a; Mousavi and Gigerenzer 2017).

---

[20] See http://www.dfg.de/en/research_funding/programmes/excellence_initiative/index.html.

In this paper, we propose to conceptualize the use of bibliometrics in research evaluation in terms of fast-and-frugal heuristics. Bibliometrics are frequently used in different evaluative contexts. They reduce the information for decision-making to publication and citation numbers. More generally speaking, this reduction may be ecologically rational in environments where active researchers are publishing researchers and where every new line of work has to be framed in terms of past research (by citing the corresponding publications). These environments can be found in most disciplines. In section "Bibliometrics-based heuristics (BBHs) illustrated: One-reason decision making in research evaluation", we gave examples of how bibliometrics can be conceptualized in terms of one-reason heuristics. However, this conceptualization can be seen as a first very rough attempt only.

In psychology (and beyond), a line of systematic research of heuristics has emerged with numerous publications, including many journal articles, foundational books (e.g., Gigerenzer et al. 1999; Hertwig et al. 2013; Todd et al. 2012), and literature that makes the fast-and-frugal heuristics research program easily accessible to practitioners and a (non-scienfic) audience (e.g., Gigerenzer 2007, 2014). Gigerenzer and Brighton (2009) present "ten well-studied heuristics for which there is evidence that they are in the adaptive toolbox of humans" (p. 130). In coining the notion "homo heuristicus" (e.g., p. 134), those authors formulate a "vision of human nature based on an adaptive toolbox of heuristics rather than on traits, attitudes, preferences, and similar internal explanations" (p. 134).

Based on the results from research on heuristics, we encourage studying the use of bibliometrics in research evaluation under the umbrella of the fast-and-frugal heuristics framework. This does not mean that the use of bibliometrics is reduced to only one heuristic model. Instead, it should be the objective to develop an adaptive toolbox including a collection of heuristics to have a coordinated set of BBH available for specific evaluation environments (Marewski and Bornmann 2019). This objective is in accordance with a recent call by Waltman and van Eck (2016) for contextualized scientometric analyses "which is based on the principles of context, simplicity, and diversity" (p. 542). Every heuristic should be tuned to specific environments and designed for specific evaluation tasks (Bornmann and Marx 2018).

Note that specifying the toolbox of BBHs does *not* necessarily entail coming up with an unlimited (or overwhelmingly large) number of heuristics—how many different tools are needed and useful would be a function of the task environments considered and the (e.g., research, applied) questions being asked. What is more, Gigerenzer and Gaissmaier's (2011) overview demonstrates how many different heuristics are crafted from common *building blocks*. As they and others point out, those building blocks allow describing the larger collection of heuristics in terms of a smaller set of more elementary and (non-arbitrarily) combinable constituents. Specifically, depending on how a smaller number of search, stopping and decision rules are matched with each other, a larger number of heuristics can be designed (Gigerenzer et al. 1999; Todd and Gigerenzer 2000). For instance, search rules that sequentially arrange predictor variables (e.g., as in Fig. 1's fast-and-frugal tree), can be complemented by even simpler search rules that contemplate predictor variables in any (e.g., random) order (see e.g., Gigerenzer et al. 2012). The search could end after a certain number (e.g., 3) of predictors has been taken into account or after a certain amount of time has elapsed (e.g., 10 min).[21] Upon the execution of those or other stopping rules, decisions could then be made by simply counting how many predictor variables suggest one option as opposed to another. This equal-weighting decision rule is built into, for instance, tallying heuristics (see e.g., Gigerenzer and Goldstein 1996; Marewski et al. 2010a, see section "How the study of heuristics can inform the study of bibliometrics"), and is helpful in many domains, ranging from medicine to financial investment and avalanche forecasting (e.g., Hafenbrädl et al. 2016). Identifying such building blocks might, too, be helpful for studying the use of BBH.

In short, in line with research on other fast-and-frugal heuristics, work on BBHs should target four key points (see section "Decision making under uncertainty: Theory and implications").

(i) *Descriptive*: The use of bibliometrics in research evaluation should be systematically studied based on the conceptual framework: Do researchers and other decision-makers in science use BBH? If so, what heuristic do they use in which context and how frequently? What are the common building blocks of those heuristics? Are there differences in the use of BBH between citizen and professional bibliometricians (Leydesdorff et al. 2016)? According to Gigerenzer and Gaissmaier (2011) "[n]umerous studies have documented systematic individual differences in the use of heuristics" (p. 459). Studies targeting those questions should work out the contents of an adaptive toolbox for research evaluation (see above). And just as research on heuristics in other areas, corresponding studies in evaluative bibliometrics should be open to the possibility that heuristics are not used at all (e.g., Glöckner et al. 2014; Jekel and Glöckner 2018b; Newell 2005), and include alternative decision mechanisms in model comparisons (Marewski et al. 2010d).[22]

(ii) *Ecological*: If BBHs are used in research evaluation, to which evaluative environments are the heuristics adapted: in which contexts does relying on a given BBH lead to accurate (and/or e.g., fast) decisions and in which environments will the BBH produce inappropriate results? When, why, and how will clear effort-accuracy trade-offs occur and when, why, and how will 'less be more'? How well do BBHs perform compared to other methods, including other indicators and more complex procedures (see section "Reasons for using bibliometrics-based heuristics (BBHs) instead of other strategies for research evaluation"). Those questions focus on the ecological rationality of BBHs and their relative performance.

(iii) *Applied*: How can research evaluation be improved? Research on the ecological rationality of heuristics can inform prescriptions and recommendations: The *International Society for Informetrics and Scientometrics* (*ISSI*)—the international asso-ciation of scholars and professionals active in the interdisciplinary study of science, science communication, and science policy[23]—could publish a list of well-studied BBH (an adaptive toolbox for the use of bibliometrics in research evaluation) alongside with the environments for which there is evidence that those heuristics lead to accurate or otherwise meaningful decisions (Bornmann and Marx 2018). Based on this adaptive toolbox, the use of BBH could be effectively taught by professional bibliometricians in workshops and courses. Moreover, from a policy-point of view, research environments could be changed so that certain heuristics work better, or alternatively, decision makers could be recommended to use other (e.g., new) heuristics than they used previously (see also Marewski and Bornmann 2019 for complementary proposals, focusing more on changing the culture of research evaluations and the statistical and field-specific expertise that ought to be required to conduct them).

---

[21] Time-based rules have been described for very different tasks in different contexts (e.g., giving-up time rules in the animal foraging literature).

[22] In addition to classic weighted-additive integration strategies and optimization approaches (e.g., utility maximization), candidate decision mechanism include, for instance, *connectionist networks* (e.g., Glöckner and Betsch 2008) and *exemplar models* (e.g., Juslin and Persson 2002) which might well predict decisions better than corresponding candidate heuristics (see, e.g., Glöckner and Betsch 2008; Heck and Erdfelder 2017).

(iv) *Methodological*: How can the usage of BBHs be studied? For one, field and laboratory studies could be used in similar ways as they have been relied upon in descriptive research on other heuristics (e.g., Brandstätter et al. 2006; Bröder and Gaissmaier 2007; Garcia-Retamero and Dhami 2009; Goldstein and Gigerenzer 2002). Those studies might not only indicate when people may rely on a given heuristic, equally important, they can also show when people do *not* rely on a given heuristic (see e.g., Bröder 2011; Johnson et al. 2008; Pohl 2006; Rieskamp and Otto 2006, for examples from other areas of research on heuristics). Secondly, the prescriptive, ecological question might be answered with mathematical analyses and computer simulation studies—again, similar to research on the ecological rationality of heuristics in other areas (e.g., Gigerenzer and Goldstein 1996; Katsikopoulos et al. 2010; Marewski and Schooler 2011; Martignon and Hoffrage 1999).

Corresponding benchmarks to gauge the performance of BBH might be machine-learning algorithms, and other advanced inference and classification machineries (e.g., Brighton 2006; Gigerenzer and Brighton 2009; Goldstein and Gigerenzer 2009). Similar to research on heuristics in sports that makes use of expert-based rankings as performance-benchmarks (e.g., Scheibehenne and Bröder 2007; Serwe and Frings 2006), other benchmarks might be, for instance, assessments made by peers. The studies by Auspurg et al. (2015), Diekmann et al. (2012), Pride and Knoth (2018), and Traag and Waltman (2019) which we discussed in section "Reasons for using bibliometrics-based heuristics (BBHs) instead of other strategies for research evaluation" are examples for comparing peer-assessments to bibliometrics.[24]

---

[23] See http://www.issi-society.org.

[24] In the introduction to this paper, we wrote that complexity can arrive at the evaluation stage when (i) many indicators are considered and (ii) evaluation procedures include numerous internal and external stages. In so doing we did not provide a precise definition of complexity. Yet, it is important to realize that the notion of complexity is, in the literature on fast-and-frugal heuristics and other computational and mathematical models, typically more formally defined, and indeed, relative model complexity can often be assessed when comparing the performance of different heuristics, optimization, and other formal models. To parallel that line of work in research evaluation, different (allegedly) more complex research evaluation procedures (e.g., multiple rounds of peer-review, composite indicators and/or combinations of those) could, ideally, be specified in terms of formal models. Then it may be possible to assess their complexity, for instance, in computer simulations as well as by taking into account drivers of model complexity, such as the number of free parameters a model requires estimating, its functional form, or the acceptable parameter space (e.g., Marewski and Olsson 2009; Marewski et al. 2010a, d; Pitt et al. 2002). To speak with Pitt et al. (2002), complexity "refers to the flexibility inherent in a model that enables it to fit diverse patterns of data" (p. 473). All by itself, considering many indicators does not necessarily imply an increased degree

## Conclusion

In research evaluation, the oldest and most frequently used method for assessing research activities is the peer review process (Bornmann 2011). This qualitative form of research evaluation can be conceived of as belonging to the class of complex (and, in the views of many: 'rational') judgment strategies. Peer review processes are frequently put in place when single contributions, such as journal submissions and grant applications must be evaluated. However, as the example in the introduction shows, the process is also used in the evaluation of many and larger units, such as universities or single researchers. Presumably, peer review processes are based on the assumption that *all* possible information (on manuscripts, researchers, applications etc.) is taken into account and then differentially weighted. Involving experts in research evaluation, most likely ensures, so the rationale goes, that assessments are based on full information (e.g., on all aspects of scientific quality). In general, this premise is only limited by reviewer tasks which exceed field-specific expertise: for instance, societal impact considerations are difficult to assess for reviewers in grant peer review procedures (Derrick and Samuel 2016).

Since the end of the 1980s, relying on mere publication and citation numbers as cue to scientific quality has become increasingly popular (Bornmann, in press). Nowadays, corresponding bibliometric indicators are often used as complement to expert peer-review; sometimes bibliometrics even fully replace expert judgments. However, bibliometrics are not yet at the center stage of a fully established profession, and, perhaps due to their openness to non-specialists and peripheral actors, a broad consensus as to which indicators should be used in what settings has yet to emerge (Jappe et al. 2018).

While all assessment methods—be they based on peer review, different bibliometric or other indicators, for that matter—have their specific advantages and disadvantages, typical arguments made in favor of resorting to bibliometrics are that:

(i) the application of bibliometrics is less costly than that of peer review (the time of experts is valuable);
(ii) large numbers of units can be evaluated and compared (it is difficult for peers to overview and assess a large number of units);
(iii) publication and citation numbers can be used to assess large units (peers have usually problems to assess large units, such as organizations and countries, because of their complexity).

---

Footnote 24 (continued)

of complexity (=flexibility) compared to a model that would consider only a few indicators. Just think of the tallying heuristic—this model takes into account several predictor variables, but it weighs them equally, requiring no parameter estimation and therefore not gaining any flexibility from estimation. Another model might take fewer predictor variables into account, but require estimating a weight for each of them, letting the model become more flexible in fitting existing data. In short, when it comes to benchmarking BBHs, a research program on BBHs may not only specify precise formal models of BBHs but also strive to formalize (allegedly more complex) approaches. That said, just as in research on other heuristics, it is, of course, possible to simply assess the performance of different BBHs against those of alternative procedures that are deemed to be relevant competitors—be it because those competitors are (i) standard procedures, and/or are (ii) particularly elaborate, time-consuming, laborious, costly, and/or resource-intensive, and/or (iii) represent expert judgments (e.g., peer review).

In research evaluation, bibliometrics and peer review are commonly used as stand-alones. The Leiden Ranking, for instance, is a sheer bibliometric ranking of universities worldwide.[25] Submissions to journals are assessed by reviewers only. However, bibliometrics and peer review can also be used in combination: in *informed peer review*, peers assess units (e.g., institutions) based on their own impressions and based on a bibliometric report. Moreover, sometimes bibliometric indicators are considered together with other indicators. Examples are university rankings (e.g., the *Academic Ranking of World Universities*[26]) which are mostly based on bibliometric measures, but additionally draw on other measures (e.g., number of Nobel laureates) to assess research performance. Although different methods do thus exist, complex procedures are generally (i.e., in all evaluation environments) conceived of as more appropriate than simpler procedures when it comes to evaluating research activities. It is seen as imperative that a complex research process with many uncertainties is evaluated with a complex—and hence seemingly rational—procedure. A related imperative comes with the usage of indicators and statistics: numbers are often seen as key to establishing seemingly 'objective' evaluations (Marewski and Bornmann 2019).

We are aware of only a few studies investigating the use of bibliometrics in different evaluative contexts empirically. For example, Hammarfelt and Haddow (2018) examined metrics use among humanities scholars in two countries and Gunashekar et al. (2017) studied how panels at the UK National Institute for Health Research (NIHR) rely on metrics. Moed et al. (1985) produced bibliometric results for research groups in the Faculty of Medicine and the Faculty of Mathematics and Natural Sciences at the University of Leiden. The results were discussed then with researchers from the two faculties.

Importantly, this latter study points to several problems in the evaluative use of bibliometric data. Indeed, as the literature overview of de Rijcke et al. (2016) shows, much work on bibliometrics focuses on "possible effects of evaluation exercises, 'gaming' of indicators, and strategic responses by scientific communities and others to requirements in research assessments" (p. 161). Thus, the bibliometric enterprise's manipulative and negative aspects are at the forefront of scientific interest, but not the possible fruitful and advantageous usage of indicators. Moreover, there does not exist a research program studying the environments in which *specific* BBH (e.g., using bibliometric indicators $X$, $Y$, and $Z$ in judgment strategy $A$, $B$, or $C$ in an informed peer review or without peer review) are more successful than other methods (e.g., pure peer review), and reversely, when other approaches work better.

Since bibliometric indicators are—as a rule—critically assessed, but used very frequently, the gap between empirical studies and practical usage is surprising (Marewski and Bornmann 2019). How is it possible that certain bibliometric indicators, such as the Journal Impact Factor, remain instruments of research evaluations although prominent and popular declarations, such as the *Declaration on Research Assessment* (*DORA*[27]), exist against its use? What are the benefits which are gained by relying on certain indicators although justified critique exists? The frequent usage might indicate that bibliometrics are simply very attractive in certain evaluative environments (e.g., because they are easy to use, i.e., 'fast-and-frugal'; for a complementary, more historical/societal interpretation, see Marewski and Bornmann 2019).

---

[25] See http://www.leidenranking.com.

[26] See http://www.shanghairanking.com.

[27] See https://sfdora.org.

To conclude, a heuristic research program on the evaluative use of bibliometrics might reveal the following: (1) BBHs can be similarly accurate as complex procedures (such as peer review) even though they are simpler. (2) The ecological rationality (e.g., accuracy) of such heuristics depends on the task environment and the goals at hand. (3) Users of bibliometrics can learn to select BBHs from an adaptive toolbox (e.g., provided by the ISSI society) as a function of the task environment and goals at hand. (4) Research evaluation comes with uncertainty which, in turn, might foster using heuristics, both from a descriptive and from a prescriptive (i.e., policy-making) point of view. (5) In addition, so one of us dares to hope (J.N.M.), such a systematic research program might also help to reflect to what extent research should be evaluated in the first place.

**Acknowledgements** Open access funding provided by Max Planck Society. We thank Katrin Auspurg, Robin Haunschild, Sven Hug, and Alexander Tekles for very helpful comments on an earlier version of this paper. We also thank Marc Jekel and another anonymous reviewer for their very thoughtful comments and feedback.